\documentclass[acmsmall, screen]{acmart}

\setcopyright{none}

\newcommand{\githublink}{https://anonymous.4open.science/r/NL2Scenic-65C8/readme.md}
\newcommand{\frameworkname}{\href{\githublink}{\textbf{\emph{NL2Scenic}}}}

\usepackage{pgfplots}
\usetikzlibrary{patterns}
\pgfplotsset{compat=1.18}

\usepackage{fvextra}

\DefineVerbatimEnvironment{prompt}{Verbatim}{
    fontsize=\small,
    numbers=left,
    numbersep=5pt,
    breaklines=true,           
    samepage=false,            
    formatcom=\relax,          
    gobble=0,                  
    xleftmargin=0pt,
    xrightmargin=0pt,
    codes={\catcode`\_=12 \catcode`\#=12}, 
    commandchars=\\\{\},        
    frame=single,
}

\usepackage{ltablex}   
\usepackage{booktabs}
\usepackage{siunitx}
\usepackage{romannum}
\usepackage{threeparttable}
\usepackage{subcaption}

\makeatletter
\@ifclassloaded{acmart}{}{%
  \providecommand{\Description}[2][]{}
}
\makeatother

\keepXColumns       

\usepackage{cleveref}
\crefformat{equation}{Equation~#2#1#3}
\Crefname{equation}{Equation}{Equations}

\usepackage{tikz}
\newcommand{\blackcircled}[1]{%
  \tikz[baseline=(char.base)]{
    \node[shape=circle, fill=black, text=white, inner sep=1pt] (char) {#1};
  }%
}

\usepackage{xcolor}
\usepackage{listings}

\definecolor{paramViolet}{RGB}{181,121,171}   
\definecolor{valueLightBlue}{RGB}{136, 172, 196}  
\definecolor{keywordBlue}{RGB}{86, 146, 179}  
\definecolor{stringOrange}{RGB}{206,127,80}     
\definecolor{sampleGreen}{RGB}{78,201,176}    
\definecolor{numberGreen}{RGB}{181, 206, 168}    
\definecolor{commentGreen}{RGB}{104, 151, 85}
\definecolor{specialYellow}{RGB}{220,220,103}

\lstdefinelanguage{Scenic}{
    sensitive=true,
    morekeywords={ego,new,Car,behavior, self},      
    morekeywords=[2]{param, model, do, try, take, interrupt, when},          
    morekeywords=[3]{Range},
    morekeywords=[4]{map, carla_map, weather, FollowLaneBehavior, target_speed, withinDistanceToObjsInLane, SetBrakeAction},
    morekeywords=[5]{FollowLaneAndStopWhenObjInLane},
    keywordstyle=\color{keywordBlue}\bfseries,
    keywordstyle=[2]\color{paramViolet}\bfseries,
    keywordstyle=[3]\color{sampleGreen}\bfseries,
    keywordstyle=[4]\color{valueLightBlue}\bfseries,
  	keywordstyle=[5]\color{specialYellow}\bfseries,
    comment=[l]{\#},
    commentstyle=\color{commentGreen}\itshape,
    stringstyle=\color{stringOrange},
    morestring=[b]",
    morestring=[b]',
}

\begin{document}

\title{David vs. Goliath: A comparative study of different-sized LLMs for code generation in the domain of automotive scenario generation}

\author{Philipp Bauerfeind}
\affiliation{%
  \institution{Clemson University}
  \city{Clemson}\state{SC}\country{USA}}
\email{pbauerf@clemson.edu}

\author{Amir Salarpour}
\affiliation{%
  \institution{Clemson University}
  \city{Clemson}\state{SC}\country{USA}}
\email{asalarp@clemson.edu}

\author{David Fernandez}
\affiliation{%
  \institution{Clemson University}
  \city{Clemson}\state{SC}\country{USA}}
\email{dferna3@clemson.edu}

\author{Pedram MohajerAnsari}
\affiliation{%
  \institution{Clemson University}
  \city{Clemson}\state{SC}\country{USA}}
\email{pmohaje@clemson.edu}

\author{Johannes Reschke}
\affiliation{%
  \institution{OTH Regensburg}
  \city{Regensburg}\country{Germany}}
\email{johannes.reschke@oth-regensburg.de}

\author{Mert D. Pesé}
\affiliation{%
  \institution{Clemson University}
  \city{Clemson}\state{SC}\country{USA}}
\email{mpese@clemson.edu}

\renewcommand{\shortauthors}{Bauerfeind et al.}

\begin{abstract}
Scenario simulation is central to testing autonomous-driving systems at scale. Scenic, a domain-specific language (DSL) paired with CARLA, enables precise, reproducible scenario specification, yet Zero-Shot/Few-Shot natural-language to Scenic (NL$\rightarrow$Scenic) generation leveraging large language models (LLMs) is hindered by scarce data, limited reproducibility, and inconsistent metrics. We present \frameworkname, an open-source dataset and framework for natural-language (NL) to Scenic generation comprising 146 NL–Scenic pairs and a difficulty-stratified 30-case test split, an \emph{Example Retriever}, and 14 prompting strategies spanning Zero-Shot (ZS), Few-Shot (FS), Chain-of-Thought (CoT), Self-Planning (SP), and Modularization-of-Thoughts (MoT). We evaluate 13 models-four proprietary (\texttt{GPT-4o}, \texttt{GPT-5}, \texttt{Claude-Sonnet-4}, \texttt{Gemini-2.5-pro}) and nine open-source code models (\texttt{Qwen2.5Coder} 0.5B–32B; \texttt{CodeLlama} 7B/13B/34B)-using text-based metrics (BLEU, ChrF, EDIT-SIM, CrystalBLEU) and execution-based metrics (compilation/generation), and validate these against an expert study with $n{=}11$ domain researchers. Edit-similarity (EDIT-SIM) exhibits the strongest correlation with human judgments; we further propose EDIT-COMP (F1 of EDIT-SIM and compilation) as a robust dataset-level proxy that improves ranking fidelity over individual metrics. Results show \texttt{GPT-4o}'s overall superiority, while \texttt{Qwen2.5Coder:14B} attains $\sim$88\% of its expert score with local deployment. Retrieval-augmented prompting, Few-Shot with Example Retriever (FSER), consistently narrows the gap for smaller models, and scaling analyses indicate diminishing returns beyond mid-size parameters, with \texttt{Qwen2.5Coder} outperforming \texttt{CodeLlama} at comparable scales. \frameworkname{} and EDIT-COMP provide a standardized, reproducible basis for evaluating Scenic code generation and suggest that mid-size open-source models are viable, cost-effective alternatives for autonomous-driving scenario programming.
\end{abstract}




\maketitle

\section{Introduction}

Autonomous driving (AD) is rapidly advancing, with companies such as Waymo~\cite{WaymoWebsite} and Lyft~\cite{lyftAutonomousRides} deploying self-driving vehicles for private transportation. As deployment scales, rigorous testing and evaluation are essential to ensure safety and reliability. Large-scale datasets such as the Waymo Open Dataset \cite{WaymoOpenDataset} and Argoverse \cite{Argoverse2} provide video and sensor data that support the development and benchmarking of AD algorithms; however, they underrepresent rare, safety-critical corner cases that are vital for robust evaluation. Because such events are difficult to capture, control, and reproduce in the real world, synthetic scenario simulations have become indispensable for controlled and repeatable testing of both safety and security aspects in AD systems \cite{SyntheticDatasets}. Domain-specific languages (DSLs), e.g., Scenic \cite{ScenicV3} and OpenSCENARIO \cite{asam_openscenario_200}, enable precise, programmatic and reproducible scenario generation at large scale. When used with CARLA \cite{CARLA}, Scenic allows the generation and execution of traffic scenarios, including those that are impractical or unsafe to record under real-world conditions. 

Prior work shows that large language models (LLMs) can translate natural-language (NL) descriptions into executable Scenic code, lowering the barrier for non-experts \cite{ScenicNL, ChatScene, Talk2Traffic, RealToSim, conversationalcodegeneration, DashcamToDrivingSimulations, Road2Code}.
Despite encouraging progress using LLMs for Scenic code generation, existing studies have three key limitations that hinder broader adoption and systematic evaluation. 

\textbf{First}, published results are difficult to reproduce, either because the frameworks rely on outdated APIs or because the frameworks themselves are not released. In addition, the absence of an unified open-source dataset prevents meaningful comparison across different studies. \textbf{Second}, systematic comparisons across model architectures are limited, with a strong focus on proprietary models, particularly \texttt{GPT-4o}. Relying solely on cloud-based models can become costly with frequent usage, whereas open-source, code-specific LLMs would allow for local deployment. \textbf{Third}, there is no standardized evaluation methodology, and existing metrics are often used without assessing their validity, which may undermine the reliability of reported results.



\begin{figure}[tbp]
    \centering
    \includegraphics[width=1\textwidth]{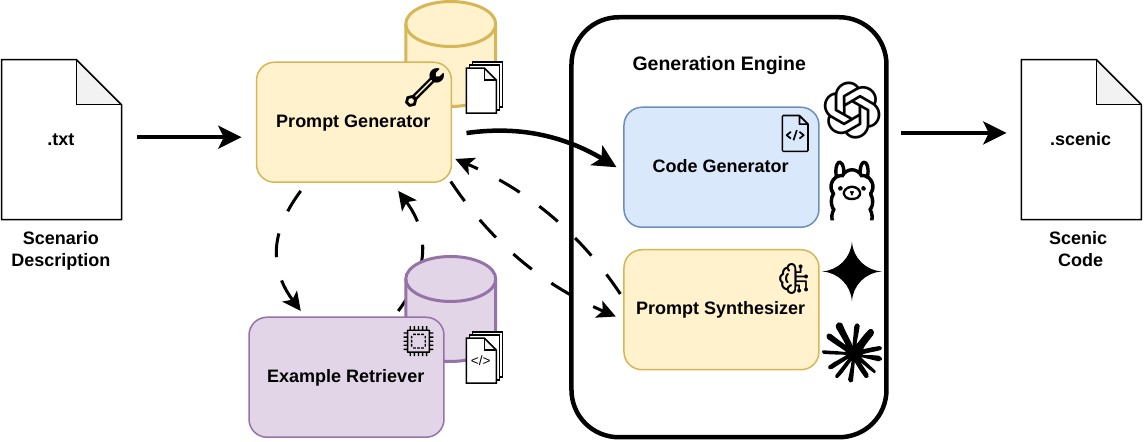}
    \caption{Architecture of \textbf{NL2Scenic} framework. The system takes a NL scenario description (.txt) as input and generates executable Scenic code (.scenic) through three main components: (1) the \emph{Prompt Generator}, which creates tailored prompts; (2) the \emph{Example Retriever}, which retrieves similar examples from a database; and (3) the \emph{Generation Engine}, which synthesizes the prompt components and invokes the LLM for code generation.}
    \label{fig:framework}
    \Description{The figure shows the overall framework: the prompt generator, the example retriever, the prompt synthesizer, and the generation engine. The generation engine comprises the code generator. Arrows indicate that the prompt generator may interact with the example retriever and the prompt synthesizer.}
\end{figure}

We introduce \frameworkname~(see \autoref{fig:framework}), an open-source dataset and framework for NL to Scenic (NL$\rightarrow$Scenic) code generation. To the best of our knowledge, it constitutes one of the largest publicly available collections of NL-Scenic paired examples, containing 146 scripts with corresponding NL description drawn from existing sources, manually crafted examples and synthetic ones. Additionally, the dataset includes a 30-case test split, with examples ranked by difficulty according to a reproducible methodology. 
The framework introduces an \emph{Example Retriever} to enhance Few-Shot prompts and provides 14 prompting strategies combining Zero-Shot, Few-Shot, Chain-of-Thought, Self-Planning, and Modularization-of-Thought variants. We evaluate 4 proprietary models (\texttt{GPT-4o}, \texttt{GPT-5}, \texttt{Claude-Sonnet-4}, \texttt{Gemini-2.5-pro}) and 9 open-source code models (\texttt{Qwen2.5Coder} \texttt{0.5B} to \texttt{32B}; \texttt{CodeLlama} \texttt{7B}/\texttt{13B}/\texttt{34B}). Performance is evaluated using pre-existing text-based metrics (BLEU, ChrF, EDIT-SIM, CrystalBLEU) and execution metrics (compilation and generation). Furthermore, we conduct a human expert study to research the validity of those metrics and propose EDIT-COMP, a composite metric defined as the F1-score of EDIT-SIM and compilation success, for Scenic code evaluation.

%

In our expert study, \texttt{GPT-4o} ranks highest, followed by \texttt{Qwen2.5Coder:14B}, which performs comparably to state-of-the-art (SOTA) commercial models. To establish a reliable evaluation methodology, we validate automatic metrics against expert judgments: EDIT-SIM shows stronger correlation with human ratings than BLEU (p$<$0.05), and EDIT-COMP further improves dataset-level ranking fidelity, providing a validated proxy for Scenic code evaluation. We find that well-designed prompting strategies, particularly Few-Shot using the \emph{Example Retriever}, enable smaller open-source models to approach the performance of proprietary alternatives. A scaling analysis suggests diminishing returns beyond a certain parameter size, with \texttt{Qwen2.5Coder} outperforming \texttt{CodeLlama} at comparable scales. 

In summary midsize open-source models can approach proprietary performance and EDIT-SIM/EDIT-COMP are valid proxies for a preliminary evaluation of Scenic code quality. 

The paper makes the following contributions:

\begin{itemize}
  \item \textbf{NL2Scenic dataset \& framework.} We release an open-source, standardized dataset (146 NL-Scenic pairs; 30-case test split) to evaluate and train NL$\rightarrow$Scenic generation, together with a comprehensive framework featuring a \emph{Example Retriever}, 14 prompting strategies and support for models across 4 distinct API platforms.
  \item \textbf{Comprehensive, model-agnostic evaluation.} We compare 4 proprietary SOTA and 9 smaller open-source code models under a unified setup, tracking text-based and execution-based performance. To our knowledge this represents the most thorough evaluation of NL$\rightarrow$Scenic generation to date.
  \item \textbf{Metric validation \& composite score.} We validate existing text- and execution-based metrics by performing a human expert study on 5 unique models. Based on our results we propose the use of EDIT-SIM/EDIT-COMP to make future evaluations more reliable.
\end{itemize}

\noindent \textbf{Availability.} We release code, data, and scripts under \texttt{MIT License} at \url{\githublink}.

\section{Related Work}
\label{sec:related}

Generating executable AD scenarios from NL combined two areas: using LLMs to generate code and programmatic scenario DSLs. Beyond Scenic~\cite{ScenicV3}, widely used formats include OpenSCENARIO and OpenDRIVE for scenario exchange and road networks, and the CommonRoad ecosystem for motion-planning benchmarks~\cite{asam_openscenario_200,CommonRoad}. We focus on NL{\textrightarrow}Scenic pipelines and relate them to adjacent DSL efforts and code-generation evaluation.

\noindent \textbf{Prompting-based NL{\textrightarrow}Scenic.}
ScenicNL~\cite{ScenicNL} combines Tree-of-Thought~\cite{TreeOfThought}, Few-Shot~\cite{LLMFewShot}, RAG~\cite{RAG}, and HyDE~\cite{HyDE} in a multi-turn strategy to generate safety-critical scenarios from NL descriptions. Applied to California DMV reports~\cite{CaDMV}, the authors report 90\% syntactic correctness. The pipeline relies on outdated APIs, making reproducabily difficult.

\noindent \textbf{Retrieval/assembly pipelines.}
ChatScene~\cite{ChatScene} decomposes NL descriptions into default settings, behaviors, geometry, and spawn positions, retrieves code snippets via embedding-based search, and assembles them into CARLA-executable Scenic scripts~\cite{CARLA}. The released scenarios use Scenic v2 syntax, leading to compatibility issues with the current release.

\noindent \textbf{Planning and fine-tuning.}
Xu~\cite{RealToSim} recreates CISS crash scenarios~\cite{CISS} and compares Zero-Shot, Few-Shot, ScenicNL, and Chain-of-Thought paired with Few-Shot~\cite{CoT}. On 100 cases, Chain-of-Thought with Few-Shot attains a compilation rate of 90\%, exceeding ScenicNL and Few-Shot ($\sim$80\%), as well as Zero-Shot (9\%). Generation rates, the fraction of compilable scripts that produce a valid CARLA simulation, are considerably lower. Strategies like self-debugging~\cite{SelfDebug} and map replacement boost generation rates by roughly 2\%. A fine-tuned \texttt{Qwen2.5Coder:1.5B}~\cite{Qwen2.5Coder} reaches 99.9\% compilation and 58.7\% generation. Semantic alignment is evaluated with ROUGE-L~\cite{ROUGE} over behavior sequences.

\noindent \textbf{Multimodal inputs (video, speech, sketch).}
Miao \emph{et al.}~\cite{DashcamToDrivingSimulations} introduce ScriptGPT (video{\textrightarrow}Scenic via \texttt{GPT-4o}~\cite{GPT4o}) with iterative refinement guided by a 10-category similarity assessment; refinement takes $\sim$1.5 minutes per scenario and yields 64\% successful generations on 50 videos. Talk2Traffic~\cite{Talk2Traffic} accepts NL, speech, and sketches; inputs are translated into a YAML intermediate (map, weather/temporal conditions, entities) and then used for RAG-guided code generation. The authors report 89\% execution success versus 15\% for Zero-Shot, as defined in their paper. Road2Code~\cite{Road2Code} is a neuro-symbolic video{\textrightarrow}Scenic pipeline combining multi-object tracking, behavior-vector encoding, and program synthesis, with reasoning distilled from \texttt{GPT-4o} to a fine-tuned \texttt{Llama3.1:8B}~\cite{llama3.1}. Evaluation includes synthetic-to-synthetic pixel/perceptual metrics and mAP@0.5~\cite{SurveyOnPerformanceMetrics}, showing improved simulation fidelity; current limitations include a single vehicle class.

\noindent \textbf{Conversational code generation with retrieval.}
Rubavicius \emph{et al.}~\cite{conversationalcodegeneration} use \texttt{CodeLlama}~\cite{CodeLlama} with RAG over 105 NL–Scenic pairs (sourced/augmented from the Scenic library~\cite{ScenicV3}) and compare against \texttt{Mistral}~\cite{mistral7b} and \texttt{Gemma}~\cite{gemma}. Text similarity (BLEU~\cite{BLEU}, ROUGE-L~\cite{ROUGE}) with leave-one-out validation~\cite{LeaveOneOut} indicates gains from RAG, code-specialized models, and human-in-the-loop refinement.

\noindent \textbf{Complementary (non-LLM) scenario generation.}
Orthogonal to NL-conditioned generation, optimization and falsification methods (e.g., counterexample-guided falsification, importance sampling, adversarial RL) search for failure cases under formal objectives or temporal-logic constraints and often integrate with Scenic-like DSLs via simulator-in-the-loop evaluation. We reference these as complementary approaches rather than empirical baselines in our study.

\noindent \textbf{Practical considerations: maps, assets, and reproducibility.}
Scenario outcomes depend on map assets and simulator versions; mixing synthetic CARLA maps with city-style layouts or OSM-derived scenes can change geometry and asset identifiers, affecting spawn feasibility and behavior scripts. To control for these factors, our evaluation pins environment versions (CARLA build, Python API), normalizes asset names when needed, and documents map replacement where applicable. We also publish prompts and post-processing scripts to support reproducibility.

\noindent \textbf{Positioning.}
Across these lines of work, three limitations recur: (i) limited cross-study comparabilty and difficulties in reproducing results,(ii) a predominant focus on proprietary models with little exploration of open-source alternatives, and (iii) inconsistent evaluation metrics that further hinder comparability. We address these gaps through three key contributions. \textbf{First}, we publish our open-source and standardized dataset, as well as our framework. \textbf{Second}, we evaluate 13 distinct models combined with 14 different prompting strategies, encompassing both proprietary and open-source LLMs (e.g., \texttt{Qwen2.5Coder}, \texttt{CodeLlama}). \textbf{Third}, we conduct an expert study with 11 domain experts to validate text- and execution-based metrics by measuring their correlation with human judgment, thereby improving the reliability of Scenic code evaluation. Our ultimate goal is a standardized, reproducible methodology for evaluating Scenic code generation. 

\section{Background}

\subsection{Scenic Programming Language and CARLA Simulator}

\begin{figure}[tbp]
    \centering
    \includegraphics[width=0.85\textwidth]{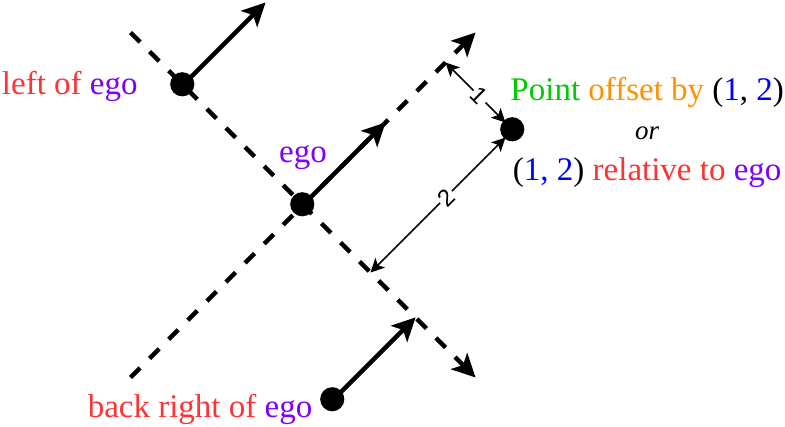}
    \caption{Scenic spatial operators for defining object relationships, including directional positioning (e.g., \emph{left of}, \emph{back right of}), point-based offsets, and relative coordinates. Adapted from~\cite{ScenicBookChapter}.}
    \label{fig:scenic-operators}
    \Description{Scenic operators (e.g., “left of”, “back right of ego”) specify spatial relationships between objects.}
\end{figure}

Scenic is a probabilistic programming language for specifying scenarios to train, test, and debug machine learning (ML) systems~\cite{ScenicBookChapter}. As ML increasingly underpins safety-critical applications, the demand for diverse, high-quality data grows, while real-world collection remains costly and resource-intensive. Synthetic data from precisely defined Scenic scenarios offer a scalable and controllable alternative.
Scenic defines scenarios as distributions over scenes comprising the spatial configuration of objects and the temporal behavior of dynamic agents~\cite{ScenicBookChapter}. It integrates with multiple simulators across domains (e.g., Webots~\cite{Webots}, X-Plane~\cite{Xplane}); in this work, we focus on automotive scenarios using the CARLA simulator. Each Scenic script includes an \emph{ego} object representing the scenario’s point of view. While Scenic’s syntax resembles Python, it adds operators that concisely express spatial relationships (see \autoref{fig:scenic-operators}).

\emph{Behaviors} define how an agent interacts with the scene. In addition to a variety of prebuilt \emph{behaviors} (e.g., \texttt{FollowLaneBehavior()}, \texttt{DriveAvoidingCollisions()}, \texttt{LaneChangeBehavior()}), Scenic also supports custom \emph{behaviors}. These can incorporate prebuilt ones or be constructed from more fine-grained \emph{actions} combined with conditional execution, as illustrated in \autoref{lst:behavior}.

\begin{lstlisting}[language=Scenic, caption=Custom behavior that follows lane and brakes when in-lane object is within specified safety distance., label={lst:behavior}]
behavior FollowLaneAndStopWhenObjInLane(speed=5, distance=10):
    try:
        # follow the lane
        do FollowLaneBehavior(target_speed=speed)
    interrupt when withinDistanceToObjsInLane(self, distance):
        # brake with full intensity when too near any object
        take SetBrakeAction(1.0)        
\end{lstlisting}

\emph{Actions} directly manipulate low-level control (e.g., brake, throttle, steering) and serve as building blocks for higher-level \emph{behaviors}. Available behaviors, actions, and other aspects (e.g., weather presets, supported object classes, vehicle \emph{blueprints}) can vary across simulators in the same domain.


\begin{figure}[t]
\centering
\includegraphics[width=1\textwidth]{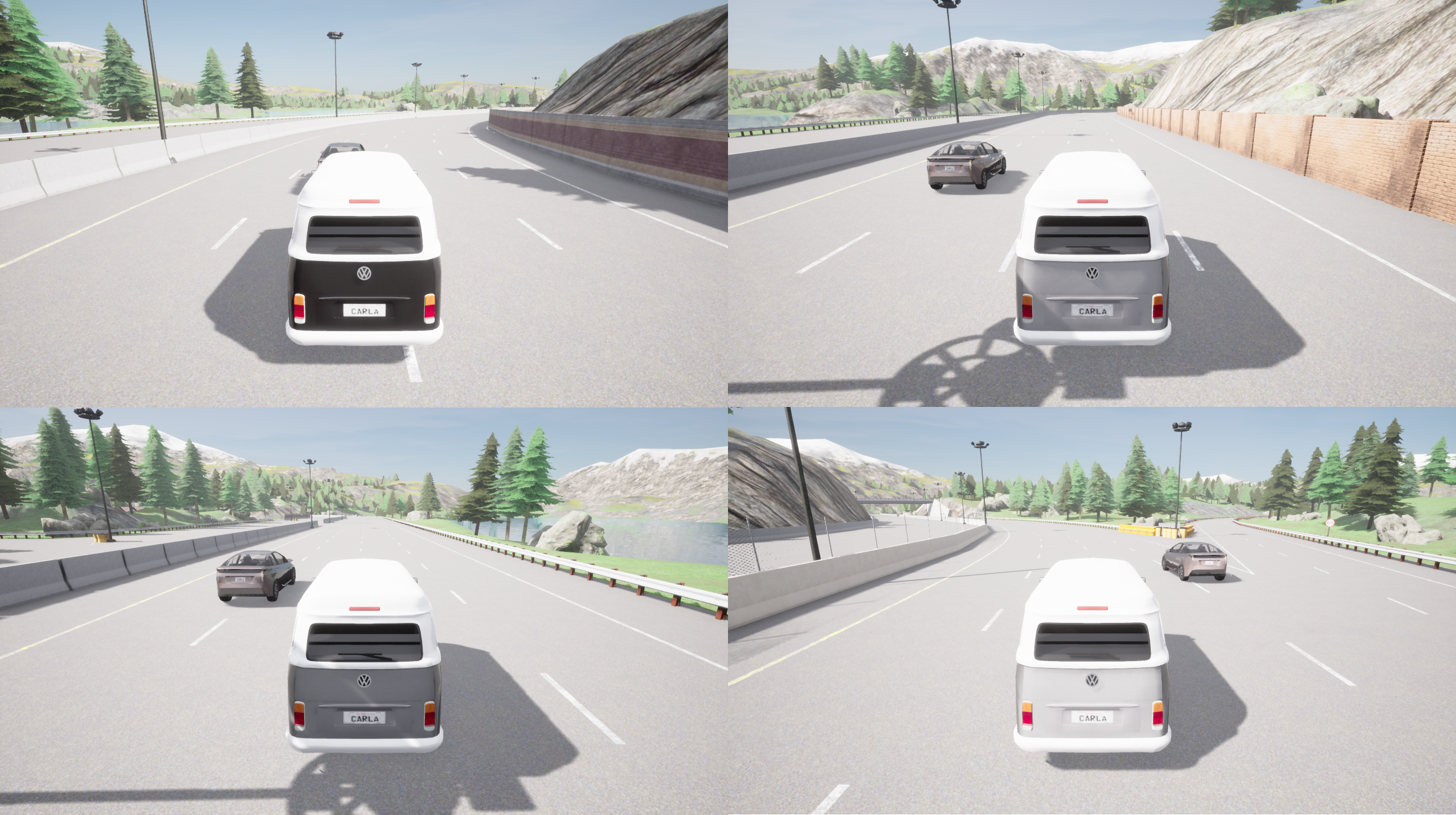}
\begin{lstlisting}[language=Scenic]
Town = 'Town04'
param map = localPath(f'../../assets/maps/CARLA/{Town}.xodr')
param carla_map = Town
model (*@scenic.simulators.carla.model@*)

param weather = 'ClearNoon'

EGO_MODEL = 'vehicle.volkswagen.t2'
OTHER_MODEL = 'vehicle.toyota.prius'

ego = new Car,
    with blueprint EGO_MODEL

c = new Car at ego offset by Range(-5, 5) @ Range(7, 12),
    with blueprint OTHER_MODEL,
    with color Color.withBytes([187, 162, 157])
\end{lstlisting}
\caption{Top: four CARLA simulations generated from single Scenic script. Bottom: corresponding Scenic script.}
\label{fig:simulation}
\Description{The figure shows an example Scenic script, as well as 4 different corresponding CARLA simulations. While all simulations were generated using the same Scenic script the appearance differs. Showcasing the probabilistic nature of Scenic.}
\end{figure}

\autoref{fig:simulation} illustrates Scenic's probabilistic scenario generation with a two-vehicle scene: a Volkswagen T2 \emph{ego} and a Toyota Prius. Although specific models are set (lines 8–9), Scenic can sample models, colors, positions, and other attributes from distributions when unspecified~\cite{ScenicBookChapter}. The Toyota is placed laterally between 5  meters to the left/right from the ego and longitudinally between 7 to 12 meter ahead of the ego (line 14).Additional specifications include the Toyota’s color (line 16) and the weather preset (line 6). The resulting simulations vary in placement and appearance due to dynamic sampling. This example is static; dynamic scenarios can attach behaviors (e.g., \texttt{FollowLaneAndStopWhenObjInLane()}) via the \texttt{with <behavior>} clause.


\subsection{Prompting Strategies}
We evaluate multiple prompting strategies for generating Scenic code and assess whether certain strategies favor particular model families.

\emph{Zero-Shot} asks the model to perform the task using only the task description, without labeled examples~\cite{PretrainPromptPredict}.
\emph{Few-Shot} augments the prompt with input–output examples to align the model to the task~\cite{LLMFewShot}.
\emph{Chain-of-Thought} decomposes the task into intermediate reasoning steps that guide code generation~\cite{CoT}.
\emph{Self-Planning} first produces a numbered plan, then leads code generation using that plan~\cite{Self-planning}.
\emph{Modularization-of-Thoughts } builds a Multilayer Reasoning Graph that structures the problem into different sublayers of abstraction prior to code generation~\cite{MoT}.






\subsection{Evaluation Metrics}


To facilitate meaningful evaluation of different code generation methods, we employ both widely-used metrics and those that have demonstrated superior performance in assessing code quality.

\noindent \textbf{BLEU.}
Among the most popular metrics for automatic evaluation of machine translation and code generation \cite{CrystalBLEU} is BLEU~\cite{BLEU}. It was designed to overcome the bottleneck of manual evaluation and operates on the modified n-gram precision $p_n$ computed for a candidate $c$ given one or more reference sequences $r$. \autoref{eq:modifiedngramprecision} shows the calculation of the modified n-gram precision.

\begin{equation}
 Count_{clip} = \min(Count_c, Count_r)
\label{eq:countclip}
\end{equation}

\begin{equation}
p_n = 
\frac{
\sum_{n\text{-gram} \in C} Count_{clip}(n\text{-gram})
}{
\sum_{n\text{-gram}' \in C'} Count(n\text{-gram}')
}
\label{eq:modifiedngramprecision}
\end{equation}

To compute it, one first counts the maximum number of times an n-gram occurs in the reference $Count_r$. Next, the number of occurrences of that n-gram in the candidate $Count_c$ is clipped by this maximum (\autoref{eq:countclip}). Dividing the clipped n-gram count, $Count_{clip}$, by the total number of n-grams in the candidate yields the modified n-gram precision, regarding a single sentence. 

\begin{equation}
BP =
\begin{cases}
1 &  \text{if } len(c) > len(r) \\
e^{1-len(r)/len(c)} & \text{if } len(c) \le len(r)
\end{cases}
\label{eq:brevitypenalty}
\end{equation}

\begin{equation}
BLEU = BP \cdot \exp\left(\sum_{n=1}^{N} w_n \log p_n \right)
\label{eq:bleu}
\end{equation}

The modified n-gram precision indirectly penalizes if the candidate is longer than the reference; furthermore, BLEU introduces a brevity penalty factor $BP$. Finally, the BLEU score can be calculated as shown in \autoref{eq:bleu}, considering n-grams of a length up to $N$ with positive weights $w_n$ summing up to one. The original paper proposes $N=4$ and $w_n=1/N$; this study adopts these standard values. While BLEU aligns well with human judgment in machine translation \cite{BLEU}, its correlation to evaluate code generation is lower compared to other text-based metrics \cite{OutOfTheBLEU, AligningOfflineMetricsWithHumanJudgment}. Despite this limitation, we include BLEU in our study due to its wide popularity.

\noindent \textbf{ChrF.}
While BLEU compares candidate and reference texts on a word or token level, ChrF~\cite{ChrF} operates on the character level. As shown in \autoref{eq:chrf}, it computes the harmonic mean of the character n-gram precision $ChrP$ and recall $ChrR$ \cite{SurveyOnPerformanceMetrics, MultiClassClassification} ($n \in [1, 6] \cap \mathbb{Z}$), analogous to the well-known F1-score \cite{MultiClassClassification} widely used in computer vision.

\begin{equation}
ChrF = 2 \cdot \frac{ChrP \cdot ChrR}{ChrP + ChrR}
\label{eq:chrf}
\end{equation}

Popovi\'{c} \cite{ChrF} demonstrated that ChrF, particularly its variant ChrF3, outperforms word-based metrics such as BLEU, TER~\cite{TER} or METEOR~\cite{METEOR} for machine translation evaluation. More recently, Evtikhiev \emph{et al.}\cite{OutOfTheBLEU} examined the alignment of commonly used text-based metrics, including BLEU, METEOR, ROUGE-L, and ChrF, alongside code-specific metrics such as CodeBLEU \cite{CodeBLEU} and RUBY \cite{RUBY}. Their evaluation of two Python-based datasets, CoNaLa~\cite{CoNaLa} and Card2code Hearthstone~\cite{Card2code}, showed that ChrF correlates most closely with human judgment, although it is not perfect. Given the similarity between Scenic and Python, we therefore decided to include ChrF in our study.  

\noindent \textbf{EDIT-SIM.}
Also preferable for judging the quality of generated code is the metric normalized edit-similarity (EDIT-SIM) \cite{IntelliCode}. EDIT-SIM is based on the Levenshtein distance \cite{Levenshtein}, which is the number of single-character edits required to transform a candidate into the reference \cite{IntelliCode, Levenshtein&BiologicalDatabaseResearch}. The metric is defined as one minus the Levenshtein distance between reference and candidate, normalized by the maximum length of the two code snippets, as shown in \autoref{eq:editsim}. 

\begin{equation}
\text{EDIT-SIM} = 1 - \frac{lev(c, r)}{max(len(c),len(r))}
\label{eq:editsim}
\end{equation}

Dibia \emph{et al.}~\cite{AligningOfflineMetricsWithHumanJudgment} recently studied the correlation between human judgment, BLEU, EDIT-SIM, and the widely known execution-based metric pass@k~\cite{pass@k}. Their study evaluated multiple LLMs on the Python-based HumanEval benchmark \cite{pass@k}, considering three rating factors: accuracy (whether the code is functionally equivalent to the reference), value (how useful the generated snippet is to a programmer) and effort to modify the code to be correct. The findings show that EDIT-SIM has a higher correlation with all three human ratings than BLEU, although it is outperformed by pass@k. However, both offline metrics are correlated with human judgment. Further analysis revealed that combining pass@k with EDIT-SIM showed the highest correlation in all categories. While Dibia \emph{et al.} recommend using pass@k, they suggest using EDIT-SIM as a viable alternative to overcome the limitations of execution-based metrics. 

\noindent \textbf{CrystalBLEU.}
This study also includes CrystalBLEU~\cite{CrystalBLEU}, a language-agnostic code evaluation metric that addresses BLEU's weakness to \emph{trivially shared n-grams}. Unlike CodeBLEU, which adds code-aware components (e.g., keyword weighting) \cite{CodeBLEU}, CrystalBLEU deliberately excludes the top $k$ most frequent n-grams from the score computation, as these carry little semantic meaning and can misleadingly inflate similarity between unrelated code snippets. Following the authors' recommendation, we set $k=500$ (optimal range: $100 \le k \le 1000$ for Java and C++). The authors demonstrate that CrystalBLEU achieves higher \emph{distinguishability}, the ratio of metric scores between semantically equivalent versus semantically different code pairs, than both BLEU and CodeBLEU. We include CrystalBLEU due to its superior discriminative ability and language-agnostic design.

\noindent \textbf{Other metrics.}
Beyond these text-based metrics, we report some basic execution-based metrics that have been used in previous studies, compilation rate~\cite{ScenicNL, RealToSim} and generation rate (percentage of simulations successfully generated)~\cite{RealToSim, conversationalcodegeneration}. Both metrics can be easily computed using predefined functions provided by the Scenic library. However, these metrics are prone to misleading results: a Scenic script consisting solely of comments would still be classified as syntactically correct, and a generated CARLA simulation might not correspond meaningfully to the original NL description. For this reason, we consider it misleading to rely solely on these two execution-based metrics without supporting human evaluation or text-based metrics. Where applicable, we also estimate the API cost per generated Scenic script.

\section{Dataset}

To enable meaningful evaluation and provide Few-Shot exemplars, we constructed a curated dataset. Public Scenic resources are scarce and often rely on outdated syntax, complicating cross-paper comparison. We therefore release \frameworkname, a consolidated collection with consistent syntax, metadata, and organization.


\subsection{Data Collection}

We aggregate three sources: the Scenic library~\cite{ScenicV3}, the \emph{ChatScene} dataset~\cite{ChatScene}, and additional synthetic scripts.


\noindent \textbf{Scenic library.}
We selected 44 driving-domain examples from the Scenic library (some CARLA-specific, others generic with minor edits) and normalized all scripts to a consistent section order:
\blackcircled{1} scenario description (docstring),
\blackcircled{2} map and model,
\blackcircled{3} constants,
\blackcircled{4} behaviors,
\blackcircled{5} spatial relations,
\blackcircled{6} scenario specification.
Some scripts omit sections or include additional ones; the ordering convention is applied throughout the dataset. The library also includes GTAV-oriented examples~\cite{GTAV} using \emph{gtaLib}~\cite{gtaLib}; these required substantial adaptation for CARLA due to simulator-specific classes and features. Using them as drafts, we produced 33 CARLA-compatible scripts following the same order. In total, Scenic library–derived content contributes 77 samples.

\begin{figure}[t]
    \centering
    \includegraphics[width=1\textwidth]{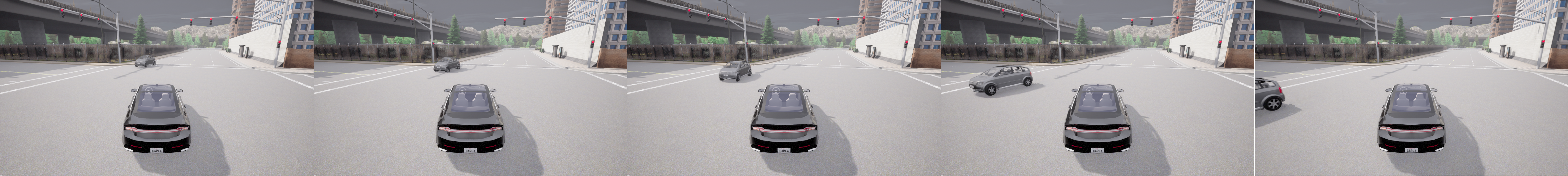}
    \includegraphics[width=1\textwidth]{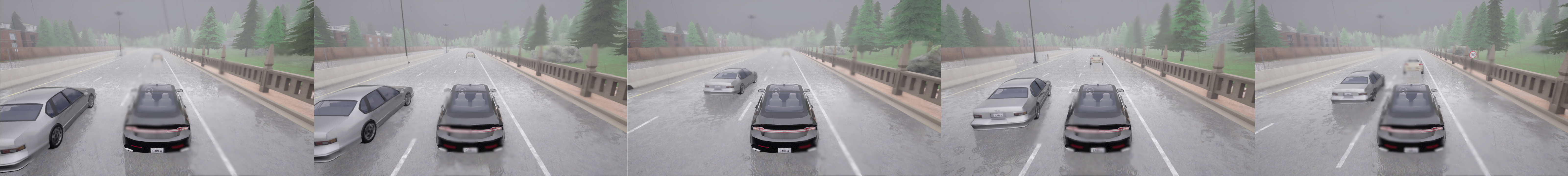}
    \caption{Example CARLA renderings from \frameworkname{} Scenic scripts. The figure shows two dinamic scenarios of the dataset. In the first scenario the ego yields to an oncoming car. The second scenario shows a multi vehicle scenario taking place on a multilane road.}
    \Description{The figure shows two scenarios. Both scenarios are visualized using 5 frames. The first scenario shows the ego vehicle performing a left turn and yielding to an oncoming car. The second scenario takes place on a multilane road: the ego follows the right most lane, while a second car overtakes the ego.}
    \label{fig:datasetexamples}
\end{figure}

\noindent \textbf{ChatScene scenarios.}
To our knowledge, \emph{ChatScene}~\cite{ChatScene} is the only other publicly available source of Scenic scenarios targeting challenging AD cases. However, the code uses Scenic v2 syntax, often misaligns with its NL descriptions, and omits ego behaviors (controlled by ML in the original study). We updated the code to current syntax, corrected NL–simulation mismatches (by editing descriptions or rewriting scenarios), and manually specified ego behaviors to match the intended descriptions. This yields 40 scripts (examples in \autoref{fig:datasetexamples}).


\noindent \textbf{Synthetic augmentation.}
To cover underrepresented CARLA attributes (weather, vehicle models/appearances, and agent classes), we generated 29 scripts from a parameterized Scenic template (see \autoref{lst:scenictemplate} in  \autoref{sec:scenictemplate}). A Python utility replaces \texttt{<keyword>} placeholders with values sampled from predefined distributions, producing valid, diverse configurations. Across all sources, \frameworkname{} comprises 146 Scenic scripts, each paired with an NL description.


\subsection{Classification \& Split}

To enable a systematic categorization of scenarios, which can support downstream model evaluation, we introduced a scoring system. Each Scenic script was assigned a score between 0 and 100, reflecting the estimated difficulty of replicating the scenario. We analyzed the curated dataset to identify indicators within the Scenic source code that could determine whether a script should be classified as \emph{Easy} or \emph{Hard}. To ensure that scores could be computed efficiently, we selected indicators that can be automatically extracted from Scenic code. Specifically, each indicator can be identified using a Python script that searches for relevant keywords or patterns in the code. The following indicators were identified, during the manual generation of Scenic scripts for the dataset: 

\begin{itemize}
  \item \textbf{Lines of Code (LoC):} Complex scenarios generally result in more lines of code.\footnote{LoC excludes comment-only lines and blank lines.}
  \item \textbf{Custom Behaviors:} Some scenarios define new behaviors, that are generally harder to reproduce than prebuilt ones.
  \item \textbf{Sub-Behaviors:} Behavioral complexity is often reflected by the use of multiple sub-behaviors.
  \item \textbf{Actions:} Complex behaviors typically involve a larger number of low-level actions.
  \item \textbf{PID Controllers (PIDs):} Highly complex behaviors may require explicit control of agents using PID controllers.
  \item \textbf{Static Agents:} more static entities increase spatial constraints.
  \item \textbf{Dynamic Agents:} Increases the number of spatial relationships within a scenario.
  \item \textbf{Requirements:} Can be difficult to formulate and introduce additional constraints. 
\end{itemize}



\begin{table}[t]
\caption{Dataset difficulty indicators (summary over 146 scripts) and indicator weights.}
\label{tab:indicators}
\centering
\begin{tabular}{lccccccc}
\toprule
\textbf{Indicator}        & \textbf{Minimum} & \textbf{Average} & \textbf{Maximum} & $\mathbf{q}_{25}$ & \textbf{Median} & $\mathbf{q}_{75}$ & \textbf{Weight} \\
\midrule
Lines of Code    & 5   & 32.062 & 86 & 12 & 33.5 & 46 & 35\% \\
Custom Behaviors          & 0   & 1.205  & 4  & 0  & 1    & 2  & 15\% \\
Sub-Behaviors             & 0   & 1.682  & 9  & 0  & 1    & 3  & 5\%  \\
Actions                   & 0   & 1.062  & 8  & 0  & 1    & 2  & 5\%  \\
PIDs                      & 0   & 0.062  & 2  & 0  & 0    & 0  & 15\% \\
Static Agents             & 0   & 1.11   & 4  & 0  & 1    & 2  & 7.5\%\\
Dynamic Agents            & 0   & 1.171  & 4  & 0  & 2    & 2  & 12.5\%\\
Requirements              & 0   & 1.137  & 5  & 0  & 1    & 2  & 5\%  \\
\bottomrule
\end{tabular}
\end{table}

Next, we collected data on these indicators for all 146 Scenic scripts (see \autoref{tab:indicators}) using a Python script that counts the occurences of each indicator. For example, to determine the number of static/dynamic agents, the script searches for the keyword \emph{new} and checks whether it is followed by a \texttt{with <behavior>} clause. If so, the agent is classified as dynamic; otherwise, it is considered static. To create a final weighted average score, each script was assigned a normalized score between 0 and 100 for each indicator. Specifically, if a value reached a score of $q_{75} + 0.5 \cdot IQR$ and above, the script received a score of 100 for that indicator. Analogously, if the value was $q_{25} - 0.5 \cdot IQR$ or less, a score of 0 was assigned -- in cases where this threshold produced negative values, 0 was used as the lower bound. For the \emph{PIDs} indicator, where $q_{25}$ and $q_{75}$ coincided, the minimum and maximum values were used instead to normalize the scores. The final scenario score was computed as a weighted average of the indicator scores, with the weights chosen heuristically (see \autoref{tab:indicators}). The highest weight was assigned to the LoC, excluding the commentary lines. Overall, scores ranged from 5.28 for the lowest scoring script. 



Consequently, the dataset was divided into three equally sized categories: \emph{Easy}, \emph{Medium} and \emph{Hard}. From each category, 10 samples were selected to construct a test dataset. The first sample in each category was chosen at random, while subsequent samples were selected by computing embeddings of the NL description using a T5-based embedding model \cite{Sentence-T5} and iteratively identifying the most dissimilar description within the remaining pool based on cosine similarity. The remaining scenarios were reserved for prompt-engineering.

\section{Scenic Code Generation}

Building on our dataset, we designed a framework (see \autoref{fig:framework}) that generates Scenic source code from NL descriptions using LLMs. It supports both proprietary and local open-source models and comprises three components: the \emph{Generation Engine}, \emph{Example Retriever}, and \emph{Prompt Generator}. Together, these modules translate NL descriptions into Scenic scripts.

\subsection{Generation Engine}

The \emph{Generation Engine} wraps multiple APIs, providing access to diverse LLMs (proprietary and open-source). It currently supports the OpenAI~\cite{openaiPlatform}, Google~\cite{googleGeminiDeveloper}, and Anthropic~\cite{anthropicAPI} platforms, as well as local execution via Ollama~\cite{ollamaWebsite}. New platforms can be added through a thin adapter that initializes credentials and normalizes request/response formats.

Beyond serving as a wrapper, the engine provides two functions: (i) direct Scenic generation from NL prompts and (ii) multistage prompting, enabling intermediate reasoning (e.g., plans or MLRs) that improves the final code-generation prompt.



\begin{figure*}[t]
    \centering
    \begin{subfigure}[t]{0.54\textwidth}
        \centering
        \includegraphics[width=\textwidth]{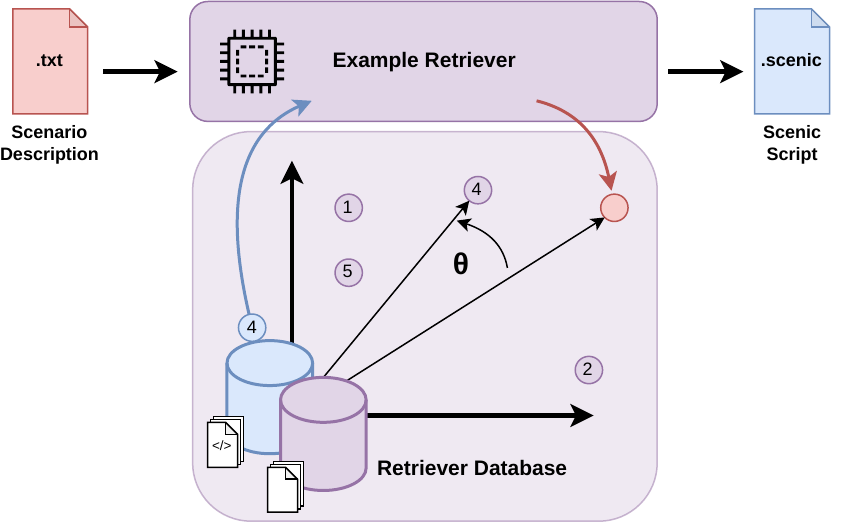}
        \caption{Example Retriever and Retriever Database. The module uses embedding-based cosine similarity to fetch relevant NL–Scenic pairs for Few-Shot prompting.}
        \label{fig:exampleretriever}
        \Description{The figure demonstrates the functionality of the example retriever. An embedding model computes the input scenario description into embeddings, which are compared to the embeddings of the retriever database. Based on cosine similarity the example retriever outputs the most similar Scenic files within the database.}
    \end{subfigure}
    \hfill
    \begin{subfigure}[t]{0.42\textwidth}
        \centering
        \includegraphics[width=\textwidth]{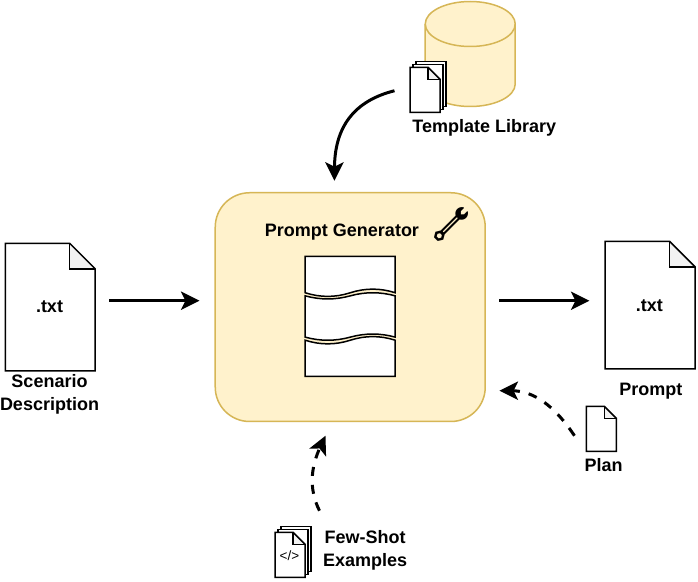}
        \caption{Prompt Generator combining scenario descriptions, templates, and optional Few-Shot examples.}
        \label{fig:promptgenerator}
        \Description{The figure shows the prompt generator module. The prompt generator has two main inputs: the scenario description and the prompt template. Optionally the prompt generator also uses Few-Shot examples or an implementation plan. Based on the chosen prompting technique the generator crafts an output prompt which is handed to the code generator.}
    \end{subfigure}
    \caption{Overview of the Example Retriever (left) and the Prompt Generator (right).}
    \label{fig:retriever_prompt_overview}
\end{figure*}

\subsection{Example Retriever}

The \emph{Example Retriever} (see \autoref{fig:exampleretriever}) is built on the \emph{all-MiniLM-L6-v2} encoder~\cite{all-Mini-L6-v2} to enhance Few-Shot performance for Scenic code generation. Although developed independently, it follows the same retrieval-augmented generation principles as prior work~\cite{conversationalcodegeneration}. The retriever has access to the \emph{Retriever Database} storing NL descriptions (violet in \autoref{fig:exampleretriever}) paired with their Scenic scripts (blue). The database is implemented as a local folder structure for easy extensibility.


At initialization, the retriever computes embeddings for all database NL descriptions, forming a local vector database. Given a new NL description, it retrieves the top-$k$ entries using cosine similarity. The paired Scenic scripts are then passed to the \emph{Prompt Generator} (see \autoref{fig:framework}).
By default we use $k {=} 3$ and index only the training split, excluding the target script to prevent leakage.


\subsection{Prompt Generator}

The \emph{Prompt Generator} (see \autoref{fig:promptgenerator}) allows users to combine each model with a variety of prompting techniques. It supports the following base strategies, which are combined or extended:


\begin{itemize}
  \item \textbf{Zero-Shot (ZS)}: The model receives only the NL description and the output format.
  \item \textbf{Few-Shot (FS)}: Expands ZS with NL-Scenic pairs.
  \item \textbf{Chain-of-Thought (CoT)}: Includes a step-by-step reasoning plan for the LLM, adding knowledge about Scenic and the CARLA simulator. This prompt was based on the prompt proposed by Xu \cite{RealToSim} and was expanded by adding more details about Scenic and CARLA. 
  \item \textbf{Self-Planning (SP)}: The model first generates a numbered implementation plan from the NL description, which is then included in the final prompt for Scenic code generation.
  \item \textbf{Modularization-of-Thought (MoT)}: The model generates a Multilayer Reasoning Graph (MLR) that divides the scenario implementation into layers of abstraction. The final prompt uses the MLR to guide code generation. 
\end{itemize}

The framework provides 14 prompting techniques (see \autoref{tab:promptingtechniques}). These techniques are combinations or variants of the base strategies and may utilize the \emph{Example Retriever} module to improve Few-Shot performance. To generate a prompt for a given strategy, the \emph{Prompt Generator} selects one of 12 templates and populates it with the required content, including the NL description, Few-Shot examples, an implementation plan, or an MLR. An example FSER prompt is shown in \autoref{sec:exampleprompt_fser}.

\begin{table}[t]
\caption{Prompting Techniques Overview (Note: fixed Examples are incorporated within the prompt, $k$ means the number of examples can be adapted and \emph{retrieved} Examples are chosen by \emph{Example Retriever}).}
\label{tab:promptingtechniques}
\resizebox{\textwidth}{!}{
\begin{tabular}{lll}
\hline
\textbf{Prompting Technique} & \multicolumn{1}{c}{\textbf{Planning Prompt}} & \textbf{Generation Prompt}                                                   \\ \hline
ZS        & \multicolumn{1}{c}{-}   & Task                                                           \\
FS        & \multicolumn{1}{c}{-}   & Task + $k=3$ Examples                                          \\
FSER      & \multicolumn{1}{c}{-}   & Task + $k=3$ related Examples                                  \\
CoT       & \multicolumn{1}{c}{-}   & Task + Reasoning Steps + Scenic Documentation                  \\
CoT-FS    & \multicolumn{1}{c}{-}   & Task + Reasoning Steps + Scenic Documentation + $k=3$ Examples \\
CoT-FSER                     & \multicolumn{1}{c}{-}                        & Task + Reasoning Steps + Scenic Documentation + $k=3$ retrieved Examples     \\
SP-ZS     & Task                    & Task + Implementation Plan + Scenic Documentation              \\
SP-FS                        & Task + 3 fixed Examples                      & Task + Implementation Plan + Scenic Documentation + $k=3$ retrieved Examples \\
SP-FS-ZS  & see SP-FS               & see SP-ZS                                                      \\
SP-ZS-FS  & see SP-ZS               & see SP-FS                                                      \\
MoT-ZS    & Task                    & Task + MLR + Scenic Documentation                              \\
MoT-FS    & Task + 3 fixed Examples & Task + MLR + Scenic Documentation + $k=3$ retrieved Examples   \\
MoT-FS-ZS & see MoT-FS              & see MoT-ZS                                                     \\
MoT-ZS-FS & see MoT-ZS              & see MoT-FS                                                     \\ \hline
\end{tabular}}
\end{table}

\section{Study Design}

\subsection{Objectives}

We systematically evaluate LLMs for Scenic code generation with a pre-specified ultimate objective.

\begin{itemize}
  \item \textbf{Model Performance.} Do models produce compilable and semantic related Scenic code, and can smaller open-source LLMs achieve SOTA performance?
  \item \textbf{Prompting Strategies.} Which prompting techniques are most effective across model sizes, and do certain strategies favor large or small models?
  \item \textbf{Metric Validity.} To what extent do automatic metrics (e.g., BLEU, ChrF, CrystalBLEU, EDIT-SIM, Compilation/Generation) reflect expert judgments of Scenic code quality? We assess alignment at both dataset and file levels via correlation tests.
\end{itemize}

\subsection{Factors and Conditions} 

\noindent \textbf{Models.}
We evaluated a diverse set of LLMs, spanning proprietary SOTA models and smaller, non-proprietary models that can be run locally. This reflects two common usage scenarios: (i) leveraging cloud-based commercial models without specialized hardware, and (ii) deploying smaller open-source models locally, which requires sufficient computing resources. All models were tested with all prompting strategies in our framework. For multistage prompting techniques, we used the same \emph{base} model for all stages.

\noindent \textbf{Proprietary Models.}
Proprietary models were accessed via commercial APIs and do not require specialized hardware. We evaluated three major platforms: OpenAI (\texttt{GPT-4o}, \texttt{GPT-5}), Anthropic (\texttt{Claude-Sonnet-4}), and Google (\texttt{Gemini-2.5-pro}). \texttt{GPT-4o} was included due to its established use in Scenic code generation, while \texttt{GPT-5} offers enhanced reasoning capabilities.

\noindent \textbf{Non-proprietary Models.}
Non-proprietary models were run locally using the Ollama framework. These open-source alternatives are well suited for downstream fine-tuning. Because Scenic is closely related to Python, we focused on code-specialized models fine-tuned for programming tasks, expecting this to translate to improved Scenic generation. We evaluated two families:

\begin{itemize}
    \item \textbf{\texttt{Qwen2.5Coder}}: six models ranging from 0.5B to 32B parameters, with strong performance on code generation benchmarks such as HumanEval \cite{huggingfaceCodeModels}.
    \item \textbf{\texttt{CodeLlama}}: three models (7B, 13B, 34B) available on Ollama, size-comparable to selected \texttt{Qwen2.5Coder} variants, enabling a comparison of model size effects within a code specific context. 
\end{itemize}

\subsection{Metrics}
\label{sec:methodologymetrics}

To assess model performance, we used text-based, execution-based, and composite metrics. Text-based metrics capture similarity between generated code and reference Scenic scripts. We also evaluated syntactic validity and executability and cost efficiency. Where possible, evaluations used standardized libraries for reproducibility. The following metrics were applied:

\begin{itemize}
    \item \textbf{BLEU}: computed with the \emph{NLTK} library \cite{NLTK}.
    \item \textbf{ChrF}: computed with the \emph{NLTK} library \cite{NLTK}.
    \item \textbf{EDIT-SIM}: Levenshtein distance via the \emph{python-Levenshtein} library \cite{LevenshteinLib} and cosequent compuation of EDIT-SIM using the standard formula (\autoref{eq:editsim}).
    \item \textbf{CrystalBLEU}: official implementation \cite{GitHubCrystalBLEU}, excluding the 500 most frequent n-grams computed from the 116 samples in the \emph{Retriever Database}.
    \item \textbf{Compilation Rate}: syntactic correctness determined by parsing generated Scenic code with the Scenic library; scripts that failed to compile were counted as incorrect.
    \item \textbf{Generation Rate}: assessed via the Scenic API to check whether a generated script can produce a valid CARLA simulation (runtime errors not considered).
    \item \textbf{Combined Metrics}: we also report simple combinations of the above metrics.
    \item \textbf{API Cost}: estimated from input/output token counts using model-specific tools.
\end{itemize}

\subsection{Expert Analysis}
\label{sec:expertanalysis}

To complement automatic evaluation, we conducted a human assessment with 11 domain experts, all of whom currently conduct or have previously conducted research in the automotive domain. Participants reported their experience with scenario simulators, including CARLA, and with the Scenic programming language. Among the 11 participants, 9 had prior experience with scenario simulators, 7 had specifically worked with CARLA, and 5 had previously used Scenic. 

\noindent \textbf{Models.}
Five model variants, as well as ground-truth references, were evaluated. We included three proprietary models and two sizes from the \texttt{Qwen2.5Coder} family, using each model’s best-performing prompting strategy from a preliminary automatic evaluation:

\begin{itemize}
    \item \texttt{Gemini-2.5-pro-FSER}
    \item \texttt{Claude-Sonnet-4-FSER}
    \item \texttt{GPT-4o-CoT-FSER} 
    \item \texttt{Qwen2.5Coder:1.5B-FSER}
    \item \texttt{Qwen2.5Coder:14B-FSER}
\end{itemize}

\noindent \textbf{Survey.}
Before the structured survey, we performed a brief qualitative review of generated scenarios to highlight characteristic strengths and weaknesses of each model and to provide context for the subsequent human ratings. We examined the same scenarios later used in the survey.
For each of the 30 NL descriptions, participants were shown:

\begin{itemize}
    \item a reference CARLA simulation and the ground-truth Scenic code
    \item a video of each \emph{successfully generated} CARLA simulation with the corresponding Scenic code.
\end{itemize}

\begin{figure}
    \centering
    \includegraphics[width=0.65\textwidth]{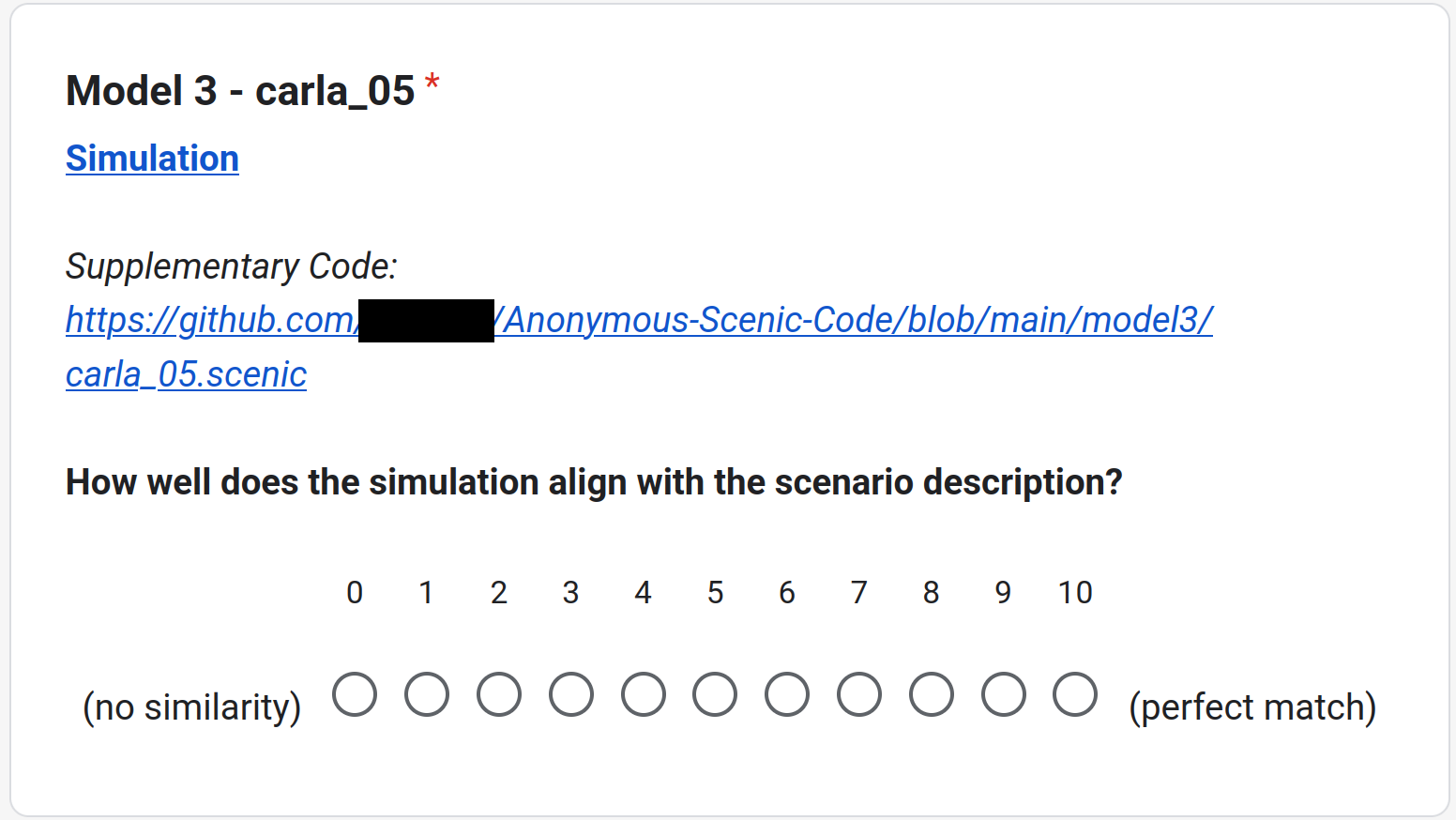}
    \caption{Screenshot of the survey: raters were given a link to the corresponding CARLA simulation, as well as the code used to generate the scenario.}
    \label{fig:surveyscreenshot}
    \Description{The figure shows a screenshot of the survey. 
    It displays an anonymized model name and scenario name, 
    two clickable links leading to the simulation and the Scenic code, 
    and a scale from 0 to 10 that the participant can click to provide a rating.}
\end{figure}

\noindent \textbf{Data Processing and Rater Reliability.}
To reduce the impact of extreme values, we applied outlier normalization (Winsorization) per question. Let $q_{25}$ and $q_{75}$ be the first and third quartiles and $IQR = q_{75} - q_{25}$. We set $T_{\text{upper}} = q_{75} + 1.5 \cdot IQR$ and $T_{\text{lower}} = q_{25} - 1.5 \cdot IQR$ and normalized any value outside this interval to the nearest boundary. Of the 1{,}220 data points, 6.82\% were clipped.

We assessed reliability using Cronbach’s alpha \cite{CronbachsAlpha}, reporting $\alpha_{\text{prenorm}} = 0.895$ before and $\alpha_{\text{norm}} = 0.865$ after normalization. Both values fall within the commonly accepted range of 0.70–0.95 \cite{CronbachsAlpha}, indicating strong internal consistency.

\noindent \textbf{Analysis.}
We compared human ratings with metrics from \autoref{sec:methodologymetrics} to assess whether text-based scores reflect perceived scenario quality. First, at the \emph{dataset level}, we compared each model’s overall expert score,rescaled to 0–100 from the mean rating across the 30 scenarios, with the model’s average metric scores computed over the test set. We evaluated significance using pairwise Williams tests \cite{WilliamsTest} among metrics. Next, at the \emph{file level}, we compared metric scores for each Scenic script with the average expert score for the corresponding generated simulation(s). We again used Williams tests among metrics and performed bootstrap resampling \cite{BootsTrap} to assess robustness.

\subsection{Text-based Evaluation}
Guided by the expert analysis, we evaluated models and prompting strategies on the test set using the validated metrics. Each model generated Scenic scripts for identical inputs, and performance was measured by computing file-level metrics and averaging them across the 30 test cases. We ranked models using each model’s optimal prompting strategy and compared the open-source families (\texttt{Qwen2.5Coder} and \texttt{CodeLlama}) to examine scaling behavior with parameter size.

\section{Evaluation}
\label{sec:evaluation}

\subsection{Experimental Setup}

All evaluations and code generation were performed on an x86-64 machine running Ubuntu 22.04.5 LTS (Linux 6.8.0-78-generic), equipped with an Intel(R) Core(TM) i3-14100 CPU and an NVIDIA GeForce RTX 3090 (GA102) GPU and 62GB of RAM. To execute Scenarios were generated using CARLA 0.9.15 and Unreal Engine 4.26.

\subsection{Expert Analysis}
As outlined in the previous section, an expert analysis was conducted to ensure a meaningful comparison of SOTA LLMs against open-source alternatives.

\begin{figure}[t]
    \centering
    \includegraphics[width=1\textwidth]{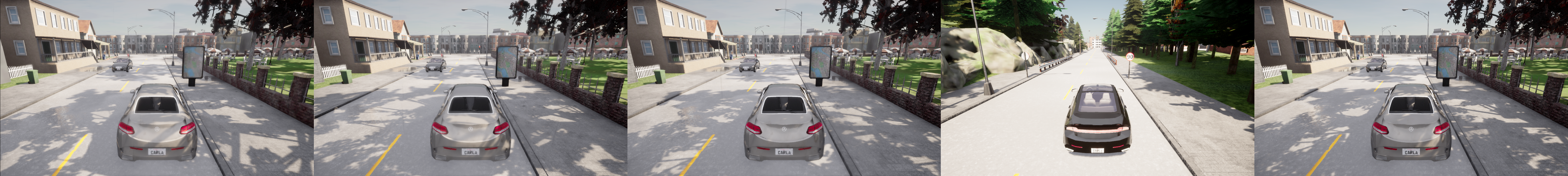}
    \caption{Side by side comparison: "The ego vehicle, a silver Mercedes Coupe is placed at (x: 41.390, y: -257.460) on map Town02. The other car, a Lincoln MKZ 2017, is positioned at (x: 45.590, y: -271.510). It's raining lightly and it is noon." (from left to right: \texttt{Gemini-2.5-pro}, \texttt{Claude-Sonnet-4}, \texttt{GPT-4o}, \texttt{Qwen2.5Coder:1.5B} and \texttt{Qwen2.5Coder:14B}).}
    \label{fig:sidebyside}
    \Description{The figure shows multiple simulations side by side. All simulations appear to be identical, except for the simulation generated by the Qwen2.5Coder:1.5B model which has no similarity to the others.}
\end{figure}

\begin{figure}[t]
    \centering
    \includegraphics[width=1\textwidth]{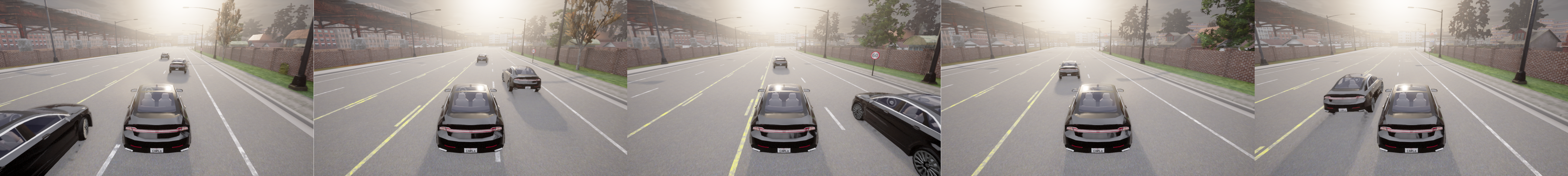}
    \caption{Example simulation generated by \texttt{Qwen2.5Coder:14B}: "Ego vehicle performs multiple lane changes to bypass three slow adversary vehicle".}
    \label{fig:simulationexample}
    \Description{The figure shows a simulation generated by the Qwen2.5Coder:14B model. The ego vehicle overtakes multiple cars.}
\end{figure}

Initially, we performed a qualitative analysis based on model simulations before conducting a larger-scale survey to examine the differences between the chosen models. \texttt{Gemini-2.5-pro} was able to generate only 11 simulations, as described in \autoref{sec:expertanalysis}. Although the quality of these simulations is high, the model mostly is able to recreate scenarios categorized as \emph{Easy}, and therefore mostly static. \texttt{Claude-Sonnet-4}, in contrast, generated 15 scenarios spanning all difficulty levels, closely resembling the corresponding NL descriptions. \texttt{GPT-4o} produced five more scenarios than Claude, effectively doubling the number of scenarios for the \emph{Hard} category. Overall, the scenarios produced by \texttt{GPT-4o} have only minor flaws. \texttt{Qwen2.5Coder:1.5B} produced 21 simulations that often deviate from the NL descriptions by omitting key elements or introducing unintended ones. In some cases, the deviations were minor, while in others the generated scenarios did not resemble the intended description; in some cases, the model failed to produce \emph{Easy} scenarios that other models could generate (see \autoref{fig:sidebyside}). Finally, \texttt{Qwen2.5Coder:14B} generated 23 valid simulations across all difficulty levels, with overall high quality. In particular, one simulation even exceeded the corresponding ground-truth simulation in fidelity (see \autoref{fig:simulationexample}). 

\begin{figure}[t]
\centering
\begin{tikzpicture}[scale=0.8, transform shape]
\begin{axis}[
    ybar,
    bar width=15pt,
    ymin=0, ymax=70,
    enlarge x limits=0.2,
    ylabel={Normalized Score},
    xtick={0,1,2,3,4},
    xticklabels={
        Gemini-2.5-pro-FSER,
        Claude-Sonnet-4-FSER,
        GPT-4o-CoT-FSER,
        Qwen2.5Coder:14B-FSER,
        Qwen2.5Coder:1.5B-FSER
    },
    x tick label style={rotate=45, anchor=east, font=\bfseries},
    yticklabel style={font=\bfseries},
    nodes near coords,
    every node near coord/.append style={font=\bfseries, anchor=south, yshift=2pt, color=black},
    bar shift=0pt,
    enlarge x limits=0.3,
]

\addplot[ybar, draw=red, fill=none, pattern=north east lines, pattern color=red] coordinates { (0,29.6) };
\addplot[ybar, draw=red, fill=none, pattern=north east lines, pattern color=red] coordinates { (1,44.6) };
\addplot[ybar, draw=red, fill=none, pattern=north east lines, pattern color=red] coordinates { (2,52.8) };
\addplot[ybar, draw=blue, fill=none, pattern=north east lines, pattern color=blue] coordinates { (3,46.7) };
\addplot[ybar, draw=blue, fill=none, pattern=north east lines, pattern color=blue] coordinates { (4,37.6) };

\end{axis}
\end{tikzpicture}
\caption{Expert evaluation scores for five LLM variants on 30 test scenarios (red: proprietary; blue: open-source).}
\Description{Expert evaluation scores for five LLM variants on 30 test scenarios (red: proprietary; blue: open-source). Scores averaged across successfully generated scenarios only. The figure shows a bar plot of all 5 models, that were evaluated during the expert analysis. The bar plots demonstrate the final values of the expert analysis normalized to a range from 0 to 100. \texttt{GPT-4o} achieved the highest value of 52.8, followed by \texttt{Qwen2.5Coder:14B} with 46.7, \texttt{Claude-Sonnet-4} with 44.6, \texttt{Qwen2.5Coder:1.5B} with 37.6 and finally \texttt{Gemini-2.5-pro} with 29.6.}\label{fig:expertanalysis}
\end{figure}
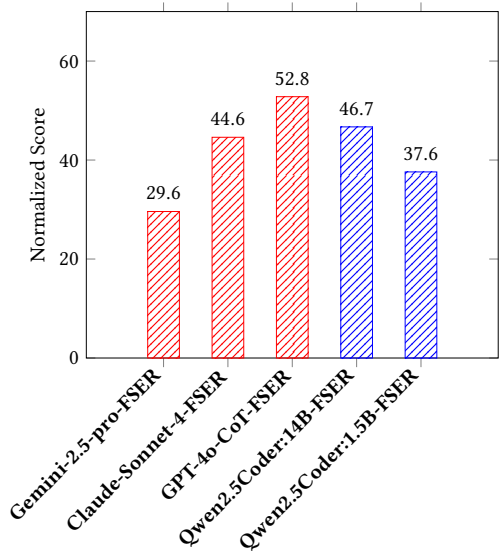

\autoref{fig:expertanalysis} shows the final results of the expert analysis: 
\texttt{GPT-4o} was rated best, followed by \texttt{Qwen2.5Coder:14B}, \texttt{Claude-Sonnet-4}, and \texttt{Gemini-2.5-pro}. \texttt{GPT-4o} achieves a top score of 52.8 compared to 93.8 for the ground-truth reference simulations. Additionally, we combined the scores of all five models for each of the three difficulty levels.  With five models, 10 test-cases per difficulty, and a maximum score of 10 per scenario the highest possible combined score per category is 500. The combined scores are 337 for \emph{Easy}, 235.5 for \emph{Medium}, and 120.1 for \emph{Hard}. The results show a downward trend with rising difficulty, indicating that, on average, the models struggle with scenarios classified as \emph{Hard}, compared to scenarios classified as \emph{Easy}.

\begin{center}
\begin{tikzpicture}
\node[fill=blue!5, draw=blue!75!black, rounded corners, inner sep=2mm, align=left, text width=13 cm] {
\textbf{Takeaway 1:} \texttt{GPT-4o} ranks highest, but \texttt{Qwen2.5Coder:14B} achieves 88\% of its performance with 14B parameters and local deployment. 
};
\end{tikzpicture}
\end{center}

\subsection{Metric Validation: Correlation with Human Judgment}
\noindent \textbf{Dataset-level.}
A critical challenge in evaluating code generation is determining which metrics reliably reflect expert judgment. While text-based metrics like BLEU and execution-based metrics like compilation rates are widely used, their validity for DSLs like Scenic remains unexplored. Metric validation is essential: without it, researchers cannot reliably compare models or assess progress. We therefore conducted a comprehensive analysis correlating automatic metrics with expert ratings to identify which metrics best reflect human judgment of Scenic code quality.

All metrics (see \autoref{tab:metriccorrelationdataset}) show positive correlation with human perception, with EDIT-SIM demonstrating the strongest correlation. Additionally, we tested metric combinations to enhance correlation. We found that combining EDIT-SIM (scaled to 100) and the compilation rate as an F1-score (EDIT-COMP) shows superior ranking ability compared to individual metrics. 

To assess statistical significance, we performed Williams tests between all metric pairs. EDIT-SIM correlates significantly better with human judgment than BLEU, CrystalBLEU, and generation rate (p<0.05). Moreover, generation rate performs significantly worse as a proxy for human perception than BLEU (p<0.10), CrystalBLEU (p<0.05), compilation rate (p<0.10), and EDIT-COMP (p<0.05).

\begin{table}[h]
\caption{Metric correlation dataset level.}
\label{tab:metriccorrelationdataset}
\resizebox{\textwidth}{!}{%
\begin{tabular}{lccccccc}
\hline
\textbf{Correlation} & \textbf{BLEU} & \textbf{ChrF} & \textbf{EDIT-SIM} & \textbf{CrystalBLEU} & \textbf{Compilation} & \textbf{Generation} & \textbf{EDIT-COMP} \\ \hline
\textbf{Pearson}     & 0.8136        & 0.8116        & 0.8451            & 0.8233               & 0.7729               & 0.5374              & 0.8090                         \\
\textbf{Spearman}    & 0.8           & 0.7           & 0.8               & 0.8                  & 0.5303               & 0.3                 & 0.9                            \\ \hline
\end{tabular}}
\end{table}

\begin{center}
\begin{tikzpicture}
\node[fill=blue!5, draw=blue!75!black, rounded corners, inner sep=2mm, align=left, text width=13 cm] {
\textbf{Takeaway 2:} EDIT-SIM shows the strongest correlation with human judgment, significantly outperforming BLEU, CrystalBLEU, and generation rate (p<0.05). Our proposed metric EDIT-COMP (F1 of EDIT-SIM and compilation rate) further improves ranking fidelity, making it the recommended metric for dataset-level Scenic code evaluation.
};
\end{tikzpicture}
\end{center}

\noindent \textbf{File-level.}
\autoref{tab:metriccorrelationfile} shows the results of our file-based evaluation. Compared to the dataset level, correlations are much weaker while still being positively correlated with human perception. As previously, we performed a Williams test between all metrics. Based on this, CrystalBLEU is weaker correlated with human perception than BLEU (>90\% confidence) and ChrF (>95\%). Performing bootstrap resampling over 100,000 samples further shows that ChrF is the best metric in 93.28\% of the cases, followed by BLEU which is the best only 3.72\% of the time. The 95\% confidence intervals are strictly positive for ChrF [0.1398, 0.5246], EDIT-SIM [0.0384, 0.4506] and BLEU [0.033, 0.4493], while the interval for CrystalBLEU includes negative values [-0.0189, 0.391]. 

\begin{table}[h]
\caption{Metric correlation file level.}
\label{tab:metriccorrelationfile}
\begin{tabular}{lcccc}
\hline
\textbf{Correlation} & \textbf{BLEU} & \textbf{ChrF} & \textbf{EDIT-SIM} & \textbf{CrystalBLEU} \\ \hline
\emph{Pearson}              & 0.2517        & 0.341         & 0.2567            & 0.1953               \\
\emph{Spearman}             & 0.2074        & 0.3477        & 0.218             & 0.1532               \\ \hline
\end{tabular}
\end{table}

\begin{center}
\begin{tikzpicture}
\node[fill=blue!5, draw=blue!75!black, rounded corners, inner sep=2mm, align=left, text width=13 cm] {
\textbf{Takeaway 3:}  Metric correlations are substantially weaker at the file level than dataset level, with ChrF emerging as the best file-level metric (93.28\% bootstrap probability). File-level metrics have limited predictive power for individual code quality.
};
\end{tikzpicture}
\end{center}

\subsection{Text-based Evaluation}

Based on the results of our expert analysis, we concluded a larger scale automatic evaluation, ranking the models based on EDIT-COMP. The complete results of this evaluation are shown in \autoref{appendix:autoeval}. \autoref{tab:modelranking} shows the results of our evaluation. Based on these results, \texttt{GPT-4o} still seems superior to other models followed by 5 out of the 6 \texttt{Qwen2.5Coder} models. Notably, \texttt{Qwen2.5Coder:1.5B} was ranked higher than \texttt{Claude-Sonnet-4}, contrary to the expert analysis. This showcases the imperfection of this ranking system. Nevertheless, most non-proprietary models seem to outperform \texttt{Gemini-2.5-pro} and \texttt{GPT-5}. Furthermore, FSER seems to be the preferred prompting strategy with the exception of \texttt{GPT-4o}, which achieves the best results with CoT-FSER and \texttt{Qwen2.5Coder:0.5B} leveraging MoT-ZS-FS. 

\begin{center}
\begin{tikzpicture}
\node[fill=blue!5, draw=blue!75!black, rounded corners, inner sep=2mm, align=left, text width=13 cm] {
\textbf{Takeaway 4}: Retrieval-augmented prompting (FSER, CoT-FSER) enables smaller models to approach SOTA performance: \texttt{Qwen2.5Coder:7B} with FSER (EDIT-COMP: 72.5) approaches \texttt{GPT-4} with CoT-FSER (74.2).
};
\end{tikzpicture}
\end{center}

\begin{table}[t]
\caption{Model ranking based on automatic evaluations.}
\label{tab:modelranking}
\begin{tabular}{lcccc}
\hline
\textbf{Model}              & \textbf{EDIT-SIM} & \textbf{Comp. [\%]} & \textbf{Gen. [\%]} & \textbf{EDIT-COMP} \\ \hline
GPT-4o-CoT-FSER             & 0.649             & 86.67                & 70                  & 74.2216            \\
Qwen2.5Coder14B-FSER        & 0.6604            & 83.33                & 76.67               & 73.6843            \\
Qwen2.5Coder:3B-FSER        & 0.6266            & 86.67                & 80                  & 72.7348            \\
Qwen2.5Coder:7B-FSER        & 0.6636            & 80                   & 70                  & 72.5444            \\
Qwen2.5Coder:1.5B-FSER      & 0.5853            & 90                   & 83.33               & 70.9311            \\
Qwen2.5Coder:32B-FSER       & 0.6602            & 73.33                & 63.33               & 69.4833            \\
Claude-sonnet-4-FSER        & 0.6081            & 73.33                & 53.33               & 66.4857            \\
CodeLlama:34B-FSER          & 0.5909            & 73.33                & 70                  & 65.4443            \\
CodeLlama:13B-FSER          & 0.5628            & 76.67                & 63.33               & 64.9114            \\
GPT-5-FSER                  & 0.5532            & 76.67                & 66.67               & 64.2683            \\
CodeLlama:7B-FSER           & 0.4528            & 70                   & 63.33               & 54.9895            \\
Gemini-2.5-pro-FSER         & 0.4509            & 40                   & 40                  & 42.3927            \\
Qwen2.5Coder:0.5B-MoT-ZS-FS & 0.0844            & 16.67                & 16.67               & 11.2063            \\ \hline
\end{tabular}
\end{table}

\begin{table}[t]
\caption{Behavior of \texttt{GPT-4o} to different prompting techniques.}
\label{tab:GPT4o}
\resizebox{\textwidth}{!}{%
\begin{tabular}{lccccc}
\hline
\multicolumn{1}{l}{\textbf{Prompting Technique}} &
  \multicolumn{1}{l}{\textbf{EDIT-SIM}} &
  \multicolumn{1}{l}{\textbf{Compilation [\%]}} &
  \multicolumn{1}{l}{\textbf{Generation [\%]}} &
  \multicolumn{1}{l}{\textbf{EDIT-COMP}} &
  \multicolumn{1}{l}{\textbf{Cost [\$USD]}} \\ \hline
ZS        & 0.2402          & 0              & 0           & 0                & \textbf{0.00243}  \\
FS        & 0.4393          & 36.67          & 20          & 39.9730          & \textbf{0.008006} \\
\textbf{FSER}      & \textbf{0.6569} & \textbf{66.67} & \textbf{60} & \textbf{66.1764} & \textbf{0.008278} \\
CoT       & 0.2044          & 23.33          & 10          & 21.7896          & 0.016587          \\
CoT-FS    & 0.4332          & 53.33          & 26.67       & 47.8066          & 0.020222          \\
\textbf{CoT-FSER}  & \textbf{0.649}  & \textbf{86.67} & \textbf{70} & \textbf{74.2216} & 0.020327          \\
SP-ZS     & 0.2522          & 6.67           & 0           & 10.5499          & 0.016661          \\
SP-FS-ZS  & 0.2894          & 3.33           & 0           & 5.9727           & 0.023951          \\
SP-ZS-FS  & 0.6034          & \textbf{66.67} & 53.33       & 63.3473          & 0.021318          \\
SP-FS     & 0.5759          & \textbf{70}    & 53.33       & 63.1915          & 0.028083          \\
MoT-ZS    & 0.2576          & 13.33          & 10          & 17.5687          & 0.01969           \\
MoT-FS-ZS & 0.284           & 6.67           & 3.33        & 10.8029          & 0.027923          \\
MoT-ZS-FS & 0.6348          & 63.33          & \textbf{60} & 63.4049          & 0.024833          \\
\textbf{MoT-FS}    & \textbf{0.6654} & \textbf{66.67} & 53.33       & \textbf{66.6049} & 0.032529          \\ \hline
\end{tabular}}
\end{table}

\noindent \textbf{Prompting Sensitivity (GPT-4o).}
To showcase the performance of different prompting techniques we want to highlight the results of the evaluation of \texttt{GPT-4o}, as both results indicate its strong performance. \autoref{tab:GPT4o} illustrates the results for all prompting techniques for \texttt{GPT-4o}. Notably, Zero-Shot generates no executable scenarios. While the EDIT-SIM of CoT is lower than for Zero-Shot, it is able to generate scenarios for every tenth NL description. Adding Few-Shot examples and furthermore leveraging the \emph{Example Retriever} to both base strategies boosts performance significantly. Based on EDIT-COMP, the best three prompting strategies are CoT-FSER, MoT-FSER, and FSER. All strategies perform worse when lacking Few-Shot examples for the code generation. 

\noindent \textbf{Cost.}
Regarding API cost, the cheapest option is Zero-Shot, at only 0.20 US cents per generation. FSER, while slightly more expensive than Few-Shot, at 0.83 US cents significantly boosts performance. CoT-FSER, the preferred prompting strategy, delivers even better results but more than doubles the cost per generation. Notably, MoT-FS, ranked as the second-best strategy in performance, is the most cost-intensive option, at 0.32 US cents.

\noindent \textbf{Scaling Behavior.}
\autoref{fig:parametersize} shows the scaling behavior of the two open-source model families, evaluated using EDIT-SIM. In the beginning, the gain in performance of increased parameters seems to be more pronounced, while at a certain point the performance seems to get saturated. Overall, the \texttt{Qwen2.5Coder} family seems to be superior to the \texttt{CodeLlama} family.

\begin{center}
\begin{tikzpicture}
\node[fill=blue!5, draw=blue!75!black, rounded corners, inner sep=2mm, align=left, text width=13 cm] {
\textbf{Takeaway 5}: Performance initially increases with model size, but saturates beyond a certain number of parameters. 
};
\end{tikzpicture}

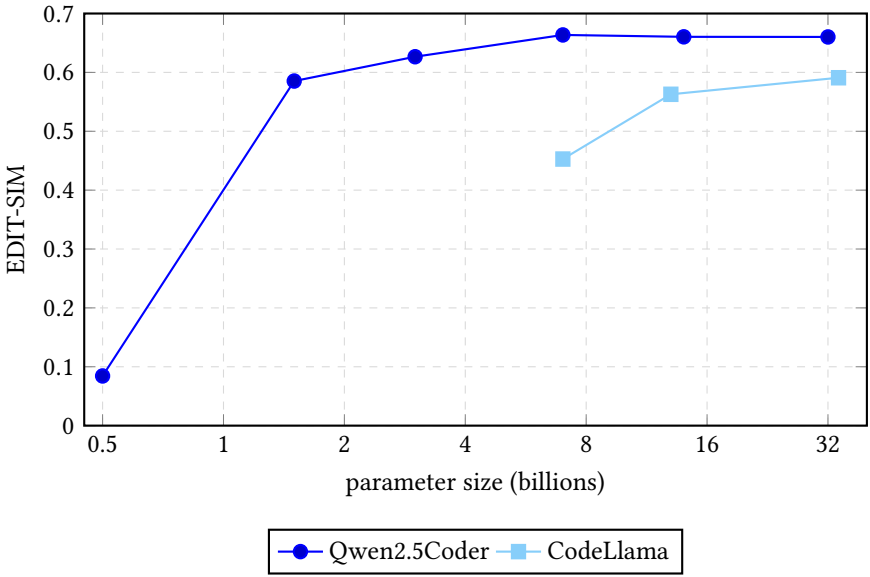
\begin{figure}[t]
\centering
\begin{tikzpicture}[scale=1, transform shape]
\begin{axis}[
    width=12cm, height=7cm,
    xlabel={parameter size (billions)},
    ylabel={EDIT-SIM},
    xmode=log, log basis x=2,
    xmin=0.45, xmax=40,
    ymin=0, ymax=0.7,
    xtick={0.5,1,2,4,8,16,32},
    xticklabels={0.5,1,2,4,8,16,32},
    ytick distance=0.1,
    grid=both,
    grid style={dashed,gray!30},
    mark size=2.5pt,
    thick,
    legend style={
        at={(0.5,-0.25)}, 
        anchor=north,
        legend columns=-1
    }
]

\addplot+[mark=*, color=blue]
coordinates {
(0.5,0.0844)
(1.5,0.5853)
(3,0.6266)
(7,0.6636)
(14,0.6604)
(32,0.6602)
};
\addlegendentry{Qwen2.5Coder}

\definecolor{lightblue}{RGB}{135,206,250} 
\addplot+[mark=square*, color=lightblue, mark options={fill=lightblue}]
coordinates {
(7,0.4528)
(13,0.5628)
(34,0.5909)
};
\addlegendentry{CodeLlama}

\end{axis}
\end{tikzpicture}
\caption{Scaling behavior of code-specialized models using EDIT-SIM metric with FSER prompting strategy.}
\label{fig:parametersize}
\Description{The figure shows the performance of the Qwen2.5Coder family and the CodeLlama with increasing parameter-size based on EDIT-SIM. At first the models performs increases with increasing parameter-size. However, beyond a certain point the performance saturates. This point seems to be at around 7 billion parameters for the Qwen2.5Coder family and 13 billion parameters for the CodeLlama family. Overall, the Qwen2.5Coder models perform better at equal parameter-size.}
\end{figure}
\end{center}

\section{Discussion}

In this study, we proposed a framework to automatically generate Scenic programs from NL descriptions: enabling the integration of multiple models from different APIs. To evaluate the effectiveness of our framework and the underlying LLMs as backbones, we performed an expert analysis, alongside an automatic evaluation that takes advantage of our newly curated dataset \frameworkname. The following discussion interprets the results from these complementary perspectives, highlighting the strengths, limitations, and potential directions for future improvements. 

The results presented in \autoref{sec:evaluation} reveal substantial differences between the LLMs evaluated. \texttt{Gemini-2.5-pro} primarily reproduces static scenarios, making it unsuitable for complex simulations. Although the quality of the scenarios it produces is high, the limited number of simulations explains the low score of our expert analysis. In contrast, \texttt{Qwen2.5Coder:1.5B} can recreate a far larger number of NL descriptions, but the scenario quality is poor. However, the model achieves a higher survey score. The larger variant \texttt{Qwen2.5Coder:14B} performs better, achieving slightly higher ratings than \texttt{Claude-Sonnet-4}, indicating that smaller code-focused models can achieve results comparable to SOTA LLMs, particularly valuable for data privacy or high-volume scenarios. 

\begin{center}
\begin{tikzpicture}
\node[fill=blue!5, draw=blue!75!black, rounded corners, inner sep=2mm, align=left, text width=13 cm] {
\textbf{Takeaway 6}: Open-source code-specialized models offer compelling cost-performance trade-offs for domain-specific generation. \texttt{Qwen2.5Coder:14B} matches \texttt{Claude-Sonnet-4}'s quality while enabling local deployment, zero API costs, and fine-tuning.
};
\end{tikzpicture}
\end{center}

Overall, \texttt{GPT-4o} remains the top-performing model, showing consistently strong results across all difficulty levels. As discussed in \autoref{sec:evaluation}, combined model scores decline with higher difficulty, confirming that our scoring method reflects the observed performance trends.

\begin{center}
\begin{tikzpicture}
\node[fill=blue!5, draw=blue!75!black, rounded corners, inner sep=2mm, align=left, text width=13 cm] {
\textbf{Takeaway 7}: \texttt{GPT-4o} remains the baseline with 20/30 successful complex scenarios and the highest expert scores (52.8).
};
\end{tikzpicture}
\end{center}

Although several models are able to generate simulations that closely match the NL descriptions, their performance remains well below our ground-truth test dataset. This highlights the need for stronger alignment with NL input as well as methods to increase the rate of successful generations. We hypothesize that both alignment and generation success could be improved through fine-tuning. In addition, approaches similar to Mia \emph{et al.} \cite{DashcamToDrivingSimulations} could be adapted to the NL setting, further improving consistency between descriptions and generated simulations.

We investigated the correlation between human perception and text-based evaluation metrics. At the dataset level, our results show that text-based metrics strongly correlate with expert judgement when applied to a dataset. EDIT-SIM is the most favorable metric for evaluating Scenic code generation, significantly surpassing BLEU and CrystalBLEU. To further address the limitations of single metrics, we propose EDIT-COMP, a combination of EDIT-SIM and the compilation rate, which demonstrates promising ranking behavior compared to other standalone metrics. At the file level, correlations are considerably weaker. ChrF performs best, significantly outperforming CrystalBLEU. However, because of the weak correlations at this granularity, we discourage the use of automatic evaluation for small datasets. Even our benchmark of 30 program description pairs would benefit from expansion to improve the reliability of automatic evaluation. 

In addition to the expert analysis, we conducted a larger-scale automatic evaluation ranking all models. Both the prompting method and the ranking order were determined using EDIT-COMP. This ranking contradicted the results of our expert analysis, highlighting the limitations of relying solely on automatic evaluation, particularly when model scores are very close. While we do not expect the rankings in \autoref{tab:modelranking} to hold under human evaluation, we argue that automatic evaluation remains useful as a preliminary proxy to narrow down the pool of models and prompting strategies for more resource-intensive manual evaluation. Especially the approach of Leung \emph{et al.} \cite{Road2Code} using computer vision to validate generations could be interesting for automatic evaluation. By creating a very specific test set of scenarios, intentionally suppressing the probabilistic nature of Scenic, models could be compared side by side leveraging computer vision metrics. According to automatic evaluation, FSER emerges as the most favorable prompting method, making it the best default choice for evaluating previously unevaluated models within our framework. Furthermore, every model listed in \autoref{tab:modelranking} benefits at some point from the \emph{Example Retriever}, underlining the importance of this module. As both expert analysis and automatic evaluation indicate the superiority of \texttt{GPT-4o}, we consider this result particularly robust. Finally, we investigated the impact of parameter size on performance using EDIT-SIM. The findings suggest that model performance saturates beyond a certain parameter threshold, implying that simply choosing the largest model does not guarantee improved results. This observation is especially relevant in the context of fine-tuning: based on our study, fine-tuning \texttt{Qwen2.5Coder:14B} appears to be the most promising direction.

\begin{center}
\begin{tikzpicture}
\node[fill=blue!5, draw=blue!75!black, rounded corners, inner sep=2mm, align=left, text width=13 cm] {
\textbf{Takeaway 8}: Text-based evaluation can be used as a proxy for preliminary results, but should be performed on a large test dataset.
};
\end{tikzpicture}
\end{center}

\section{Conclusion}

In this work, we introduced a framework for generating Scenic programs for the CARLA simulator directly from NL descriptions. Using our new dataset and framework \frameworkname, we evaluated the performance of several LLMs through expert analysis and automatic evaluation. Our results highlight the strong performance of open-source LLMs, making them a viable alternative to SOTA LLMs. At the same time, \texttt{GPT-4o} consistently outperforms all other tested models, confirming its robustness across scenario difficulties.  
We also investigated the validity of text-based metrics as proxies for human judgment. Our findings suggest that EDIT-SIM and our proposed composite metric EDIT-COMP
provide useful approximations at the dataset level. These metrics can serve as a preliminary evaluation method to narrow down the pool of candidate models before conducting more resource-intensive evaluations. Finally, the \texttt{Qwen2.5Coder} family emerges as a particularly promising direction for future work, as these models already achieve strong results without domain-specific fine-tuning. We expect that targeted fine-tuning could further boost their performance and help close the gap with larger proprietary models.

\clearpage

\bibliographystyle{ACM-Reference-Format}
\bibliography{references}

\newpage
\appendix

\section{Synthetic Data Template}
\label{sec:scenictemplate}

\begin{lstlisting}[language=Scenic, caption=Synthetic data Scenic template., label={lst:scenictemplate}]
"""Scenario Description:

The scene shows a <Color> <CarBlueprint> and a <Type> <Distance> meters
ahead in the same lane as the ego vehicle. <Weather>.

"""

#################################
# MAP AND MODEL                 #
#################################

Town = <Town>
param map = localPath(f'../../assets/maps/CARLA/{Town}.xodr')
param carla_map = Town
model (*@scenic.simulators.carla.model@*)

#################################
# CONSTANTS                     #
#################################

WEATHER_OPTIONS = <WeatherCode>
param weather = Uniform(*WEATHER_OPTIONS)

EGO_MODEL = <CarCode>

#################################
# SCENARIO SPECIFICATION        #
#################################

ego = new Car,
    with blueprint EGO_MODEL,
    with color Color.withBytes([<ColorCode>])

new <TypeCode> following roadDirection from ego for <DistanceCode>,
    with regionContainedIn ego.laneSection
\end{lstlisting}

\newpage
\section{Example Prompt}
\label{sec:exampleprompt_fser}
\begin{prompt}
Return a Scenic (probabilistic programming language) script for the CARLA simulator based on the following scenario description:

"The ego vehicle follows a road, when a pedestrian suddenly crosses the street."

The towns/maps are in the relative folder path: '../../assets/maps/CARLA/'.

Here are some examples of Scenic code and the according scenario descriptions as comment:

----------------------------------------------ExamplesBegin
-----------------------ScenicBegin

"""Scenario Description:

The ego vehicle is driving on a straight road when a pedestrian suddenly crosses from the right front and suddenly stops as the ego vehicle approaches.

"""

#################################
# MAP AND MODEL                 #
#################################

Town = 'Town05'
param map = localPath(f'../../assets/maps/CARLA/{Town}.xodr') 
param carla_map = Town
model scenic.simulators.carla.model

#################################
# CONSTANTS                     #
#################################

EGO_MODEL = "vehicle.lincoln.mkz_2017"

param OPT_EGO_SPEED = Range(1, 5)
param OPT_ADV_SPEED = Range(1, 5)
param OPT_ADV_DISTANCE = Range(15, 20)
param OPT_BRAKE_DIST = Range(6, 10)
param OPT_GEO_X_DISTANCE = Range(3, 5)
param OPT_GEO_Y_DISTANCE = Range(20, 35)

OPT_STOP_DISTANCE = 1

#################################
# AGENT BEHAVIORS               #
#################################

behavior WaitBehavior():
    while True:
        wait

behavior CrossAndStopBehavior(actor_reference, adv_speed, adv_distance, stop_reference, stop_distance):
    do CrossingBehavior(actor_reference, adv_speed, adv_distance) until (distance from self to stop_reference <= stop_distance)
    take SetWalkingSpeedAction(0)

behavior EgoBehavior():
    try:
        do FollowLaneBehavior(globalParameters.OPT_EGO_SPEED)
    interrupt when (withinDistanceToObjsInLane(self, globalParameters.OPT_BRAKE_DIST)):
        take SetThrottleAction(0)
        take SetBrakeAction(1)
        do WaitBehavior() for 5 seconds
        terminate

#################################
# SPATIAL RELATIONS             #
#################################

intersection = Uniform(*filter(lambda i: i.is4Way and not i.isSignalized, network.intersections))
egoInitLane = Uniform(*intersection.incomingLanes)
egoManeuver = Uniform(*filter(lambda m: m.type is ManeuverType.STRAIGHT, egoInitLane.maneuvers))
egoTrajectoryLine = egoInitLane.centerline + egoManeuver.connectingLane.centerline + egoManeuver.endLane.centerline

egoSpawnPt = new OrientedPoint in egoManeuver.startLane.centerline
IntSpawnPt = new OrientedPoint following egoInitLane.orientation from egoSpawnPt for globalParameters.OPT_GEO_Y_DISTANCE

#################################
# SCENARIO SPECIFICATION        #
#################################

ego = new Car at egoSpawnPt,
    with regionContainedIn None,
    with blueprint EGO_MODEL,
    with behavior EgoBehavior()

AdvAgent = new Pedestrian right of IntSpawnPt by globalParameters.OPT_GEO_X_DISTANCE,
    with heading IntSpawnPt.heading + 90 deg,  # Heading perpendicular to the road, adjusted for left crossing
    with regionContainedIn None,
    with behavior CrossAndStopBehavior(ego, globalParameters.OPT_ADV_SPEED, globalParameters.OPT_ADV_DISTANCE, egoTrajectoryLine, OPT_STOP_DISTANCE)

require 40 <= (distance to intersection) <= 60

-----------------------ScenicEnd
-----------------------ScenicBegin

"""Scenario Description:

The ego-vehicle is following a road with a parked car on the right side, next to the road. A pedestrian suddenly crosses the road from behind the parked car, forcing the ego to brake.

"""

#################################
# MAP AND MODEL                 #
#################################

Town = 'Town01'
param map = localPath(f'../../assets/maps/CARLA/{Town}.xodr')
param carla_map = Town
model scenic.domains.driving.model

#################################
# CONSTANTS                     #
#################################

PEDESTRIAN_TRIGGER_DISTANCE = 15     # Distance at which pedestrian begins to cross
BRAKE_TRIGGER_DISTANCE = 10          # Distance at which ego begins braking
EGO_TO_PARKED_CAR_MIN_DIST = 30      # Ensure ego starts far enough away
PEDESTRIAN_OFFSET = 3                # Offset for pedestrian placement ahead of parked car
PARKED_CAR_OFFSET = 1                # Offset for parked car from the curb

#################################
# AGENT BEHAVIORS               #
#################################

behavior DriveAndBrakeForPedestrians():
    try:
        do FollowLaneBehavior()
    interrupt when withinDistanceToAnyPedestrians(self, BRAKE_TRIGGER_DISTANCE):
        take SetThrottleAction(0), SetBrakeAction(1)

#PEDESTRIAN BEHAVIOR: Pedestrian crosses road when ego is near
behavior CrossRoad():
    while distance from self to ego > PEDESTRIAN_TRIGGER_DISTANCE:
        wait
    take SetWalkingDirectionAction(self.heading), SetWalkingSpeedAction(1)

#################################
# SCENARIO SPECIFICATION        #
#################################

ego = new Car with behavior DriveAndBrakeForPedestrians()

rightCurb = ego.laneGroup.curb
spot = new OrientedPoint on visible rightCurb

parkedCar = new Car right of spot by PARKED_CAR_OFFSET, with regionContainedIn None

require distance from ego to parkedCar > EGO_TO_PARKED_CAR_MIN_DIST

new Pedestrian ahead of parkedCar by PEDESTRIAN_OFFSET,
    facing 90 deg relative to parkedCar,
    with behavior CrossRoad()

terminate after 30 seconds

-----------------------ScenicEnd
-----------------------ScenicBegin

"""Scenario Description:

The ego vehicle is turning left at an intersection; the adversarial pedestrian on the right of the target lane suddenly crosses the road and stops in the middle of the road.

"""

#################################
# MAP AND MODEL                 #
#################################

Town = 'Town05'
param map = localPath(f'../../assets/maps/CARLA/{Town}.xodr') 
param carla_map = Town
model scenic.simulators.carla.model

#################################
# CONSTANTS                     #
#################################

EGO_MODEL = "vehicle.lincoln.mkz_2017"

param OPT_ADV_SPEED = Range(1, 5)
param OPT_ADV_DISTANCE = Range(15, 20)
param OPT_BRAKE_DIST = Range(6, 10)
param OPT_EGO_SPEED = Range(1, 5)

OPT_STOP_DISTANCE = 1
OPT_PARAM_LANE_WIDTH = 6

#################################
# AGENT BEHAVIORS               #
#################################

behavior WaitBehavior():
    while True:
        wait

behavior CrossAndStopBehavior(actor_reference, adv_speed, adv_distance, stop_reference, stop_distance):
    do CrossingBehavior(actor_reference, adv_speed, adv_distance) until (distance from self to stop_reference <= stop_distance)
    take SetWalkingSpeedAction(0)

behavior EgoBehavior():
    try:
        do FollowTrajectoryBehavior(globalParameters.OPT_EGO_SPEED, egoTrajectory)
    interrupt when (withinDistanceToObjsInLane(self, globalParameters.OPT_BRAKE_DIST)):
        take SetThrottleAction(0)
        take SetBrakeAction(1)
        do WaitBehavior() for 5 seconds
        abort
    terminate

#################################
# SPATIAL RELATIONS             #
#################################

intersection = Uniform(*filter(lambda i: i.is4Way or i.is3Way, network.intersections))
egoManeuver = Uniform(*filter(lambda m: m.type is ManeuverType.LEFT_TURN, intersection.maneuvers))
egoInitLane = egoManeuver.startLane
egoTrajectory = [egoInitLane, egoManeuver.connectingLane, egoManeuver.endLane]
egoTrajectoryLine = egoInitLane.centerline + egoManeuver.connectingLane.centerline + egoManeuver.endLane.centerline

egoSpawnPt = new OrientedPoint in egoInitLane.centerline
# Spawn point on the far side of the intersection, along the end lane's centerline
endLanePt = new OrientedPoint at egoManeuver.endLane.rightEdge.start,
    with heading egoInitLane.centerline.end.heading - 180 deg
pedSpawnPt = new OrientedPoint ahead of endLanePt by - OPT_PARAM_LANE_WIDTH

#################################
# SCENARIO SPECIFICATION        #
#################################

ego = new Car at egoSpawnPt,
    with regionContainedIn None,
    with blueprint EGO_MODEL,
    with behavior EgoBehavior()

AdvAgent = new Pedestrian at pedSpawnPt,
    with heading pedSpawnPt.heading,  # Perpendicular to the road, crossing the street
    with regionContainedIn None,
    with behavior CrossAndStopBehavior(ego, globalParameters.OPT_ADV_SPEED, globalParameters.OPT_ADV_DISTANCE, egoTrajectoryLine, OPT_STOP_DISTANCE)

require 40 <= (distance to intersection) <= 60

-----------------------ScenicEnd
----------------------------------------------ExamplesEnd

Important: You must only return one single coherent Scenic program in the following format:

```scenic

"""Scenario Description:

<SCENARIO_DESCRIPTION>

"""

<SCENIC_PROGRAM>

```
\end{prompt}

\newpage
\section{Complete Results of Automatic Evaluation}
\label{appendix:autoeval}

\begin{tabularx}{\textwidth}{>{\raggedright\arraybackslash}l
                             >{\centering\arraybackslash}X
                             >{\centering\arraybackslash}X
                             >{\centering\arraybackslash}X
                             >{\centering\arraybackslash}X
                             >{\centering\arraybackslash}X
                             >{\centering\arraybackslash}X
                             >{\centering\arraybackslash}X}
                             
\toprule
\multicolumn{8}{c}{\textbf{ChatGPT-4o}} \\  
\midrule
\scriptsize\textbf{Prompting Technique} & 
\scriptsize\textbf{BLEU} & 
\scriptsize\textbf{ChrF} & 
\scriptsize\textbf{EDIT-SIM} & 
\scriptsize\textbf{CrystalBLEU} & 
\scriptsize\textbf{Compilation [\%]} & 
\scriptsize\textbf{Generation [\%]} & 
\scriptsize\textbf{Cost [\$USD]} \\
\midrule
ZS & 0.1771 & 0.2885 & 0.2402 & 0.1345 & 0 & 0 & 0.00243 \\
FS & 0.4066 & 0.7045 & 0.4393 & 0.2339 & 36.67 & 20 & 0.008006 \\
FSER & 0.6148 & 0.8204 & 0.6569 & 0.4825 & 66.67 & 60 & 0.008278 \\
CoT & 0.1727 & 0.3997 & 0.2044 & 0.0915 & 23.33 & 10 & 0.016587 \\
CoT-FS & 0.3992 & 0.7030 & 0.4332 & 0.2322 & 53.33 & 26.67 & 0.020222 \\
CoT-FSER & 0.6113 & 0.8196 & 0.6490 & 0.4827 & 86.67 & 70 & 0.020327 \\
SP-ZS & 0.2185 & 0.3977 & 0.2522 & 0.1326 & 6.67 & 0 & 0.016661 \\
SP-FS-ZS & 0.2264 & 0.4059 & 0.2894 & 0.1323 & 3.33 & 0 & 0.023951 \\
SP-ZS-FS & 0.5571 & 0.7975 & 0.6034 & 0.4249 & 66.67 & 53.33 & 0.021318 \\
SP-FS & 0.5405 & 0.7913 & 0.5759 & 0.3890 & 70 & 53.33 & 0.028083 \\
MoT-ZS & 0.2081 & 0.3697 & 0.2576 & 0.1436 & 13.33 & 10 & 0.019690 \\
MoT-FS-ZS & 0.2126 & 0.3969 & 0.2840 & 0.1310 & 6.67 & 3.33 & 0.027923 \\
MoT-ZS-FS & 0.5868 & 0.7977 & 0.6348 & 0.4442 & 63.33 & 60 & 0.024833 \\
MoT-FS & 0.5479 & 0.7865 & 0.6654 & 0.4040 & 66.67 & 53.33 & 0.032529 \\

\toprule
\multicolumn{8}{c}{\textbf{GPT-5}} \\
\midrule
\scriptsize\textbf{Prompting Technique} & 
\scriptsize\textbf{BLEU} & 
\scriptsize\textbf{ChrF} & 
\scriptsize\textbf{EDIT-SIM} & 
\scriptsize\textbf{CrystalBLEU} & 
\scriptsize\textbf{Compilation [\%]} & 
\scriptsize\textbf{Generation [\%]} & 
\scriptsize\textbf{Cost [\$USD]} \\
\midrule
ZS & 0.1237 & 0.3396 & 0.1737 & 0.0842 & 0.00 & 0.00 & 0.005693 \\
FS & 0.2679 & 0.6969 & 0.3176 & 0.1363 & 50.00 & 40.00 & 0.009043 \\
FSER & 0.5077 & 0.8267 & 0.5532 & 0.3761 & 76.67 & 66.67 & 0.007903 \\
CoT & 0.1290 & 0.4273 & 0.1948 & 0.0635 & 33.33 & 3.33 & 0.014633 \\
CoT-FS & 0.2152 & 0.6799 & 0.2689 & 0.0988 & 36.67 & 33.33 & 0.015947 \\
CoT-FSER & 0.4181 & 0.7906 & 0.4779 & 0.2905 & 80.00 & 63.33 & 0.014621 \\
SP-ZS & 0.0800 & 0.4089 & 0.1270 & 0.0390 & 6.67 & 3.33 & 0.023291 \\
SP-FS-ZS & 0.0914 & 0.4173 & 0.1467 & 0.0423 & 3.33 & 3.33 & 0.022387 \\
SP-ZS-FS & 0.1513 & 0.6748 & 0.1982 & 0.0722 & 20.00 & 13.33 & 0.024654 \\
SP-FS & 0.1623 & 0.6936 & 0.2149 & 0.0746 & 10.00 & 3.33 & 0.025182 \\
MoT-ZS & 0.1153 & 0.4185 & 0.1685 & 0.0561 & 16.67 & 6.67 & 0.028818 \\
MoT-FS-ZS & 0.1143 & 0.4300 & 0.1737 & 0.0553 & 6.67 & 3.33 & 0.034010 \\
MoT-ZS-FS & 0.3295 & 0.7633 & 0.3886 & 0.2170 & 50.00 & 30.00 & 0.029811 \\
MoT-FS & 0.2667 & 0.7343 & 0.3211 & 0.1568 & 56.67 & 23.33 & 0.036298 \\

\toprule
\multicolumn{8}{c}{\textbf{Claude-Sonnet-4}} \\  
\midrule
\scriptsize\textbf{Prompting Technique} & 
\scriptsize\textbf{BLEU} & 
\scriptsize\textbf{ChrF} & 
\scriptsize\textbf{EDIT-SIM} & 
\scriptsize\textbf{CrystalBLEU} & 
\scriptsize\textbf{Compilation [\%]} & 
\scriptsize\textbf{Generation [\%]} & 
\scriptsize\textbf{Cost [\$USD]} \\
\midrule
ZS & 0.2028 & 0.3829 & 0.2261 & 0.1441 & 3.33 & 3.33 & 0.006686 \\
FS & 0.3453 & 0.7135 & 0.3870 & 0.1986 & 33.33 & 26.67 & 0.017805 \\
FSER & 0.5739 & 0.8305 & 0.6081 & 0.4409 & 73.33 & 53.33 & 0.016524 \\
CoT & 0.1966 & 0.4440 & 0.2332 & 0.1147 & 36.67 & 6.67 & 0.031061 \\
CoT-FS & 0.3249 & 0.7142 & 0.3632 & 0.1748 & 46.67 & 36.67 & 0.038392 \\
CoT-FSER & 0.5434 & 0.8182 & 0.5849 & 0.4072 & 70.00 & 53.33 & 0.036682 \\
SP-ZS & 0.1724 & 0.4454 & 0.2193 & 0.0980 & 26.67 & 6.67 & 0.031609 \\
SP-FS-ZS & 0.1819 & 0.4548 & 0.2403 & 0.1033 & 36.67 & 13.33 & 0.042690 \\
SP-ZS-FS & 0.4140 & 0.7968 & 0.4627 & 0.2785 & 56.67 & 50.00 & 0.038772 \\
SP-FS & 0.4165 & 0.7844 & 0.4640 & 0.2771 & 76.67 & 53.33 & 0.049927 \\
MoT-ZS & 0.2357 & 0.4341 & 0.2760 & 0.1401 & 26.67 & 20.00 & 0.037832 \\
MoT-FS-ZS & 0.2091 & 0.4578 & 0.2630 & 0.1192 & 30.00 & 10.00 & 0.050672 \\
MoT-ZS-FS & 0.4995 & 0.8086 & 0.5480 & 0.3441 & 56.67 & 53.33 & 0.046237 \\
MoT-FS & 0.4763 & 0.8031 & 0.5172 & 0.3396 & 60.00 & 46.67 & 0.057546 \\

\toprule
\multicolumn{8}{c}{\textbf{Gemini-2.5-pro}} \\
\midrule
\scriptsize\textbf{Prompting Technique} & 
\scriptsize\textbf{BLEU} & 
\scriptsize\textbf{ChrF} & 
\scriptsize\textbf{EDIT-SIM} & 
\scriptsize\textbf{CrystalBLEU} & 
\scriptsize\textbf{Compilation [\%]} & 
\scriptsize\textbf{Generation [\%]} & 
\scriptsize\textbf{Cost [\$USD]} \\
\midrule
ZS & 0.1149 & 0.3754 & 0.1599 & 0.0772 & 0.00 & 0.00 & 0.006589 \\
FS & 0.2234 & 0.6958 & 0.2718 & 0.1030 & 10.00 & 10.00 & 0.010194 \\
FSER & 0.4063 & 0.7892 & 0.4509 & 0.2750 & 40.00 & 40.00 & 0.009036 \\
CoT & 0.1177 & 0.4445 & 0.1714 & 0.0632 & 20.00 & 3.33 & 0.015470 \\
CoT-FS & 0.2291 & 0.6961 & 0.2763 & 0.1063 & 16.67 & 16.67 & 0.016415 \\
CoT-FSER & 0.3491 & 0.7637 & 0.3942 & 0.2248 & 43.33 & 30.00 & 0.015843 \\
SP-ZS & 0.1404 & 0.4381 & 0.1896 & 0.0802 & 0.00 & 0.00 & 0.015320 \\
SP-FS-ZS & 0.1324 & 0.4483 & 0.1864 & 0.0738 & 3.33 & 3.33 & 0.021960 \\
SP-ZS-FS & 0.2502 & 0.6939 & 0.2985 & 0.1347 & 20.00 & 16.67 & 0.018040 \\
SP-FS & 0.2361 & 0.7238 & 0.2834 & 0.1163 & 46.67 & 26.67 & 0.023943 \\
MoT-ZS & 0.1773 & 0.4292 & 0.2256 & 0.1054 & 0.00 & 0.00 & 0.023855 \\
MoT-FS-ZS & 0.1462 & 0.4415 & 0.2048 & 0.0795 & 10.00 & 6.67 & 0.026697 \\
MoT-ZS-FS & 0.3090 & 0.7192 & 0.3478 & 0.1675 & 26.67 & 16.67 & 0.026468 \\
MoT-FS & 0.2639 & 0.7116 & 0.3146 & 0.1425 & 13.33 & 6.67 & 0.028800 \\

\toprule
\multicolumn{8}{c}{\textbf{Qwen2.5Coder:0.5B}} \\
\midrule
\scriptsize\textbf{Prompting Technique} & 
\scriptsize\textbf{BLEU} & 
\scriptsize\textbf{ChrF} & 
\scriptsize\textbf{EDIT-SIM} & 
\scriptsize\textbf{CrystalBLEU} & 
\scriptsize\textbf{Compilation [\%]} & 
\scriptsize\textbf{Generation [\%]} & 
\scriptsize\textbf{Cost [\$USD]} \\
\midrule
ZS & 0.0748 & 0.2134 & 0.1152 & 0.0568 & 0.00 & 0.00 & 0.000 \\
FS & 0.1249 & 0.5303 & 0.1594 & 0.0388 & 0.00 & 0.00 & 0.000 \\
FSER & 0.1750 & 0.3621 & 0.2340 & 0.1112 & 3.33 & 3.33 & 0.000 \\
CoT & 0.0987 & 0.2117 & 0.1293 & 0.0786 & 0.00 & 0.00 & 0.000 \\
CoT-FS & 0.1444 & 0.2623 & 0.1921 & 0.1122 & 0.00 & 0.00 & 0.000 \\
CoT-FSER & 0.1506 & 0.2727 & 0.2079 & 0.1254 & 0.00 & 0.00 & 0.000 \\
SP-ZS & 0.0220 & 0.1525 & 0.0727 & 0.0158 & 3.33 & 3.33 & 0.000 \\
SP-FS-ZS & 0.0481 & 0.2159 & 0.1096 & 0.0331 & 10.00 & 10.00 & 0.000 \\
SP-ZS-FS & 0.1475 & 0.2969 & 0.2116 & 0.1044 & 0.00 & 0.00 & 0.000 \\
SP-FS & 0.0929 & 0.2062 & 0.1560 & 0.0649 & 0.00 & 0.00 & 0.000 \\
MoT-ZS & 0.0228 & 0.1353 & 0.0694 & 0.0149 & 6.67 & 6.67 & 0.000 \\
MoT-FS-ZS & 0.0156 & 0.1782 & 0.0483 & 0.0091 & 3.33 & 3.33 & 0.000 \\
MoT-ZS-FS & 0.0346 & 0.1362 & 0.0844 & 0.0248 & 16.67 & 16.67 & 0.000 \\
MoT-FS & 0.0265 & 0.1547 & 0.0560 & 0.0140 & 0.00 & 0.00 & 0.000 \\

\toprule
\multicolumn{8}{c}{\textbf{Qwen2.5Coder:1.5B}} \\
\midrule
\scriptsize\textbf{Prompting Technique} & 
\scriptsize\textbf{BLEU} & 
\scriptsize\textbf{ChrF} & 
\scriptsize\textbf{EDIT-SIM} & 
\scriptsize\textbf{CrystalBLEU} & 
\scriptsize\textbf{Compilation [\%]} & 
\scriptsize\textbf{Generation [\%]} & 
\scriptsize\textbf{Cost [\$USD]} \\
\midrule
ZS & 0.1425 & 0.2120 & 0.2111 & 0.1126 & 0.00 & 0.00 & 0.000 \\
FS & 0.2072 & 0.6227 & 0.2467 & 0.0843 & 100.00 & 100.00 & 0.000 \\
FSER & 0.5609 & 0.7840 & 0.5853 & 0.4162 & 90.00 & 83.33 & 0.000 \\
CoT & 0.1430 & 0.1886 & 0.2106 & 0.1168 & 10.00 & 0.00 & 0.000 \\
CoT-FS & 0.2579 & 0.6344 & 0.2976 & 0.1318 & 50.00 & 40.00 & 0.000 \\
CoT-FSER & 0.4613 & 0.6662 & 0.4956 & 0.3494 & 36.67 & 23.33 & 0.000 \\
SP-ZS & 0.0205 & 0.1889 & 0.0707 & 0.0106 & 3.33 & 0.00 & 0.000 \\
SP-FS-ZS & 0.0541 & 0.1743 & 0.1168 & 0.0329 & 3.33 & 0.00 & 0.000 \\
SP-ZS-FS & 0.3500 & 0.5968 & 0.3886 & 0.2059 & 46.67 & 43.33 & 0.000 \\
SP-FS & 0.2951 & 0.5370 & 0.3431 & 0.1830 & 36.67 & 30.00 & 0.000 \\
MoT-ZS & 0.0824 & 0.2074 & 0.1355 & 0.0551 & 0.00 & 0.00 & 0.000 \\
MoT-FS-ZS & 0.0544 & 0.2500 & 0.1035 & 0.0377 & 0.00 & 0.00 & 0.000 \\
MoT-ZS-FS & 0.2528 & 0.4666 & 0.3106 & 0.1576 & 43.33 & 36.67 & 0.000 \\
MoT-FS & 0.2300 & 0.4494 & 0.2688 & 0.1349 & 26.67 & 20.00 & 0.000 \\

\toprule
\multicolumn{8}{c}{\textbf{Qwen2.5Coder:3B}} \\
\midrule
\scriptsize\textbf{Prompting Technique} & 
\scriptsize\textbf{BLEU} & 
\scriptsize\textbf{ChrF} & 
\scriptsize\textbf{EDIT-SIM} & 
\scriptsize\textbf{CrystalBLEU} & 
\scriptsize\textbf{Compilation [\%]} & 
\scriptsize\textbf{Generation [\%]} & 
\scriptsize\textbf{Cost [\$USD]} \\
\midrule
ZS & 0.1506 & 0.2506 & 0.2017 & 0.1122 & 0.00 & 0.00 & 0.000 \\
FS & 0.2128 & 0.6285 & 0.2535 & 0.0893 & 93.33 & 90.00 & 0.000 \\
FSER & 0.6023 & 0.8116 & 0.6266 & 0.4683 & 86.67 & 80.00 & 0.000 \\
CoT & 0.1698 & 0.2426 & 0.2073 & 0.1253 & 0.00 & 0.00 & 0.000 \\
CoT-FS & 0.3416 & 0.6589 & 0.3898 & 0.1856 & 36.67 & 6.67 & 0.000 \\
CoT-FSER & 0.5186 & 0.7153 & 0.5670 & 0.4280 & 50.00 & 40.00 & 0.000 \\
SP-ZS & 0.0219 & 0.2398 & 0.0688 & 0.0133 & 0.00 & 0.00 & 0.000 \\
SP-FS-ZS & 0.0696 & 0.2556 & 0.1439 & 0.0431 & 0.00 & 0.00 & 0.000 \\
SP-ZS-FS & 0.3230 & 0.6469 & 0.3608 & 0.1973 & 30.00 & 20.00 & 0.000 \\
SP-FS & 0.3561 & 0.6241 & 0.4137 & 0.2043 & 36.67 & 26.67 & 0.000 \\
MoT-ZS & 0.0363 & 0.2181 & 0.0863 & 0.0221 & 0.00 & 0.00 & 0.000 \\
MoT-FS-ZS & 0.0360 & 0.2753 & 0.1087 & 0.0213 & 0.00 & 0.00 & 0.000 \\
MoT-ZS-FS & 0.2588 & 0.5304 & 0.2836 & 0.1500 & 20.00 & 13.33 & 0.000 \\
MoT-FS & 0.3067 & 0.5979 & 0.3394 & 0.1677 & 20.00 & 13.33 & 0.000 \\

\toprule
\multicolumn{8}{c}{\textbf{Qwen2.5Coder:7B}} \\
\midrule
\scriptsize\textbf{Prompting Technique} & 
\scriptsize\textbf{BLEU} & 
\scriptsize\textbf{ChrF} & 
\scriptsize\textbf{EDIT-SIM} & 
\scriptsize\textbf{CrystalBLEU} & 
\scriptsize\textbf{Compilation [\%]} & 
\scriptsize\textbf{Generation [\%]} & 
\scriptsize\textbf{Cost [\$USD]} \\
\midrule
ZS & 0.1301 & 0.2600 & 0.1903 & 0.0954 & 0.00 & 0.00 & 0.000 \\
FS & 0.2746 & 0.6728 & 0.3178 & 0.1342 & 63.33 & 53.33 & 0.000 \\
FSER & 0.6271 & 0.8127 & 0.6636 & 0.5064 & 80.00 & 70.00 & 0.000 \\
CoT & 0.1878 & 0.2663 & 0.2111 & 0.1401 & 0.00 & 0.00 & 0.000 \\
CoT-FS & 0.3770 & 0.7032 & 0.4091 & 0.2063 & 36.67 & 26.67 & 0.000 \\
CoT-FSER & 0.5553 & 0.7451 & 0.6049 & 0.4434 & 53.33 & 43.33 & 0.000 \\
SP-ZS & 0.0256 & 0.2365 & 0.0936 & 0.0146 & 0.00 & 0.00 & 0.000 \\
SP-FS-ZS & 0.0744 & 0.2857 & 0.1529 & 0.0357 & 0.00 & 0.00 & 0.000 \\
SP-ZS-FS & 0.4435 & 0.7147 & 0.4808 & 0.2839 & 40.00 & 33.33 & 0.000 \\
SP-FS & 0.4218 & 0.6977 & 0.4446 & 0.2583 & 30.00 & 20.00 & 0.000 \\
MoT-ZS & 0.0740 & 0.2328 & 0.1247 & 0.0465 & 0.00 & 0.00 & 0.000 \\
MoT-FS-ZS & 0.0572 & 0.2465 & 0.1212 & 0.0315 & 0.00 & 0.00 & 0.000 \\
MoT-ZS-FS & 0.3952 & 0.6678 & 0.4502 & 0.2547 & 30.00 & 23.33 & 0.000 \\
MoT-FS & 0.3751 & 0.6597 & 0.4294 & 0.2178 & 36.67 & 30.00 & 0.000 \\

\toprule
\multicolumn{8}{c}{\textbf{Qwen2.5Coder:14B}} \\
\midrule
\scriptsize\textbf{Prompting Technique} & 
\scriptsize\textbf{BLEU} & 
\scriptsize\textbf{ChrF} & 
\scriptsize\textbf{EDIT-SIM} & 
\scriptsize\textbf{CrystalBLEU} & 
\scriptsize\textbf{Compilation [\%]} & 
\scriptsize\textbf{Generation [\%]} & 
\scriptsize\textbf{Cost [\$USD]} \\
\midrule
ZS & 0.1417 & 0.2344 & 0.2066 & 0.1100 & 0.00 & 0.00 & 0.000 \\
FS & 0.3700 & 0.7023 & 0.4039 & 0.2085 & 36.67 & 16.67 & 0.000 \\
FSER & 0.6269 & 0.8313 & 0.6604 & 0.5008 & 83.33 & 76.67 & 0.000 \\
CoT & 0.1772 & 0.2249 & 0.2311 & 0.1394 & 0.00 & 0.00 & 0.000 \\
CoT-FS & 0.3637 & 0.6922 & 0.3925 & 0.1995 & 43.33 & 23.33 & 0.000 \\
CoT-FSER & 0.5897 & 0.8088 & 0.6237 & 0.4711 & 73.33 & 60.00 & 0.000 \\
SP-ZS & 0.0342 & 0.2323 & 0.0793 & 0.0119 & 0.00 & 0.00 & 0.000 \\
SP-FS-ZS & 0.0691 & 0.3129 & 0.1566 & 0.0348 & 0.00 & 0.00 & 0.000 \\
SP-ZS-FS & 0.3720 & 0.7236 & 0.3966 & 0.2219 & 36.67 & 33.33 & 0.000 \\
SP-FS & 0.4006 & 0.7281 & 0.4267 & 0.2380 & 43.33 & 26.67 & 0.000 \\
MoT-ZS & 0.0721 & 0.2137 & 0.1062 & 0.0390 & 0.00 & 0.00 & 0.000 \\
MoT-FS-ZS & 0.0926 & 0.2850 & 0.1411 & 0.0538 & 0.00 & 0.00 & 0.000 \\
MoT-ZS-FS & 0.4597 & 0.7627 & 0.4991 & 0.3061 & 73.33 & 66.67 & 0.000 \\
MoT-FS & 0.3851 & 0.7021 & 0.4140 & 0.2358 & 36.67 & 23.33 & 0.000 \\

\toprule
\multicolumn{8}{c}{\textbf{Qwen2.5Coder:32B}} \\
\midrule
\scriptsize\textbf{Prompting Technique} & 
\scriptsize\textbf{BLEU} & 
\scriptsize\textbf{ChrF} & 
\scriptsize\textbf{EDIT-SIM} & 
\scriptsize\textbf{CrystalBLEU} & 
\scriptsize\textbf{Compilation [\%]} & 
\scriptsize\textbf{Generation [\%]} & 
\scriptsize\textbf{Cost [\$USD]} \\
\midrule
ZS & 0.1669 & 0.2707 & 0.2123 & 0.1240 & 0.00 & 0.00 & 0.000 \\
FS & 0.3537 & 0.6916 & 0.3901 & 0.1999 & 20.00 & 3.33 & 0.000 \\
FSER & 0.6296 & 0.8358 & 0.6602 & 0.4984 & 73.33 & 63.33 & 0.000 \\
CoT & 0.1796 & 0.2794 & 0.1964 & 0.1246 & 0.00 & 0.00 & 0.000 \\
CoT-FS & 0.3380 & 0.6875 & 0.3698 & 0.1812 & 26.67 & 16.67 & 0.000 \\
CoT-FSER & 0.5899 & 0.8165 & 0.6131 & 0.4602 & 56.67 & 46.67 & 0.000 \\
SP-ZS & 0.0441 & 0.2758 & 0.0873 & 0.0191 & 0.00 & 0.00 & 0.000 \\
SP-FS-ZS & 0.0792 & 0.3070 & 0.1376 & 0.0345 & 0.00 & 0.00 & 0.000 \\
SP-ZS-FS & 0.3500 & 0.7482 & 0.3878 & 0.1930 & 30.00 & 26.67 & 0.000 \\
SP-FS & 0.3453 & 0.7288 & 0.3805 & 0.1746 & 3.33 & 0.00 & 0.000 \\
MoT-ZS & 0.0835 & 0.2655 & 0.1220 & 0.0459 & 0.00 & 0.00 & 0.000 \\
MoT-FS-ZS & 0.0650 & 0.2973 & 0.1208 & 0.0328 & 0.00 & 0.00 & 0.000 \\
MoT-ZS-FS & 0.4035 & 0.7503 & 0.4275 & 0.2415 & 20.00 & 16.67 & 0.000 \\
MoT-FS & 0.3552 & 0.7214 & 0.3980 & 0.2096 & 13.33 & 10.00 & 0.000 \\

\toprule
\multicolumn{8}{c}{\textbf{CodeLlama:7B}} \\
\midrule
\scriptsize\textbf{Prompting Technique} & 
\scriptsize\textbf{BLEU} & 
\scriptsize\textbf{ChrF} & 
\scriptsize\textbf{EDIT-SIM} & 
\scriptsize\textbf{CrystalBLEU} & 
\scriptsize\textbf{Compilation [\%]} & 
\scriptsize\textbf{Generation [\%]} & 
\scriptsize\textbf{Cost [\$USD]} \\
\midrule
ZS & 0.1189 & 0.2467 & 0.1684 & 0.0856 & 3.33 & 3.33 & 0.000 \\
FS & 0.2055 & 0.5399 & 0.2412 & 0.0911 & 50.00 & 50.00 & 0.000 \\
FSER & 0.4128 & 0.5863 & 0.4528 & 0.3198 & 70.00 & 63.33 & 0.000 \\
CoT & 0.1541 & 0.2466 & 0.1969 & 0.1144 & 0.00 & 0.00 & 0.000 \\
CoT-FS & 0.1996 & 0.4498 & 0.2309 & 0.1233 & 6.67 & 6.67 & 0.000 \\
CoT-FSER & 0.3820 & 0.5562 & 0.4268 & 0.2904 & 40.00 & 26.67 & 0.000 \\
SP-ZS & 0.0350 & 0.2022 & 0.0875 & 0.0180 & 0.00 & 0.00 & 0.000 \\
SP-FS-ZS & 0.0488 & 0.2428 & 0.1289 & 0.0292 & 3.33 & 3.33 & 0.000 \\
SP-ZS-FS & 0.2583 & 0.5176 & 0.3038 & 0.1390 & 23.33 & 16.67 & 0.000 \\
SP-FS & 0.2129 & 0.4779 & 0.2660 & 0.1101 & 16.67 & 3.33 & 0.000 \\
MoT-ZS & 0.1075 & 0.2233 & 0.1608 & 0.0805 & 10.00 & 10.00 & 0.000 \\
MoT-FS-ZS & 0.0663 & 0.2269 & 0.1467 & 0.0434 & 0.00 & 0.00 & 0.000 \\
MoT-ZS-FS & 0.1934 & 0.4246 & 0.2252 & 0.1226 & 6.67 & 0.00 & 0.000 \\
MoT-FS & 0.2299 & 0.4763 & 0.2935 & 0.1174 & 20.00 & 0.00 & 0.000 \\

\toprule
\multicolumn{8}{c}{\textbf{CodeLlama:13B}} \\
\midrule
\scriptsize\textbf{Prompting Technique} & 
\scriptsize\textbf{BLEU} & 
\scriptsize\textbf{ChrF} & 
\scriptsize\textbf{EDIT-SIM} & 
\scriptsize\textbf{CrystalBLEU} & 
\scriptsize\textbf{Compilation [\%]} & 
\scriptsize\textbf{Generation [\%]} & 
\scriptsize\textbf{Cost [\$USD]} \\
\midrule
ZS & 0.1474 & 0.2470 & 0.2116 & 0.1097 & 0.00 & 0.00 & 0.000 \\
FS & 0.2419 & 0.5551 & 0.2861 & 0.1301 & 60.00 & 56.67 & 0.000 \\
FSER & 0.5040 & 0.6368 & 0.5628 & 0.4011 & 76.67 & 63.33 & 0.000 \\
CoT & 0.1497 & 0.2239 & 0.2079 & 0.1128 & 0.00 & 0.00 & 0.000 \\
CoT-FS & 0.3353 & 0.6486 & 0.3751 & 0.1880 & 13.33 & 3.33 & 0.000 \\
CoT-FSER & 0.4722 & 0.6631 & 0.5012 & 0.3614 & 43.33 & 40.00 & 0.000 \\
SP-ZS & 0.0224 & 0.1859 & 0.0764 & 0.0110 & 0.00 & 0.00 & 0.000 \\
SP-FS-ZS & 0.0612 & 0.2482 & 0.1425 & 0.0365 & 0.00 & 0.00 & 0.000 \\
SP-ZS-FS & 0.2227 & 0.4903 & 0.2601 & 0.1199 & 23.33 & 16.67 & 0.000 \\
SP-FS & 0.2377 & 0.5067 & 0.2913 & 0.1260 & 23.33 & 13.33 & 0.000 \\
MoT-ZS & 0.0882 & 0.1961 & 0.1476 & 0.0626 & 3.33 & 0.00 & 0.000 \\
MoT-FS-ZS & 0.0728 & 0.2304 & 0.1383 & 0.0474 & 0.00 & 0.00 & 0.000 \\
MoT-ZS-FS & 0.3001 & 0.5188 & 0.3324 & 0.2027 & 40.00 & 33.33 & 0.000 \\
MoT-FS & 0.2687 & 0.4768 & 0.3284 & 0.1628 & 33.33 & 23.33 & 0.000 \\

\toprule
\multicolumn{8}{c}{\textbf{CodeLlama:34B}} \\
\midrule
\scriptsize\textbf{Prompting Technique} & 
\scriptsize\textbf{BLEU} & 
\scriptsize\textbf{ChrF} & 
\scriptsize\textbf{EDIT-SIM} & 
\scriptsize\textbf{CrystalBLEU} & 
\scriptsize\textbf{Compilation [\%]} & 
\scriptsize\textbf{Generation [\%]} & 
\scriptsize\textbf{Cost [\$USD]} \\
\midrule
ZS & 0.1360 & 0.2389 & 0.2011 & 0.1014 & 0.00 & 0.00 & 0.000 \\
FS & 0.2684 & 0.6644 & 0.3077 & 0.1261 & 73.33 & 66.67 & 0.000 \\
FSER & 0.5603 & 0.7573 & 0.5909 & 0.4405 & 73.33 & 70.00 & 0.000 \\
CoT & 0.1606 & 0.2461 & 0.2108 & 0.1161 & 0.00 & 0.00 & 0.000 \\
CoT-FS & 0.2777 & 0.4379 & 0.3244 & 0.1946 & 43.33 & 20.00 & 0.000 \\
CoT-FSER & 0.5340 & 0.7131 & 0.5785 & 0.4127 & 63.33 & 46.67 & 0.000 \\
SP-ZS & 0.0413 & 0.2135 & 0.1010 & 0.0139 & 0.00 & 0.00 & 0.000 \\
SP-FS-ZS & 0.0851 & 0.2610 & 0.1615 & 0.0504 & 0.00 & 0.00 & 0.000 \\
SP-ZS-FS & 0.2789 & 0.5896 & 0.3316 & 0.1524 & 26.67 & 23.33 & 0.000 \\
SP-FS & 0.3590 & 0.6385 & 0.4064 & 0.1969 & 10.00 & 10.00 & 0.000 \\
MoT-ZS & 0.1269 & 0.2607 & 0.1701 & 0.0920 & 0.00 & 0.00 & 0.000 \\
MoT-FS-ZS & 0.0953 & 0.2344 & 0.1448 & 0.0573 & 0.00 & 0.00 & 0.000 \\
MoT-ZS-FS & 0.2775 & 0.4654 & 0.3210 & 0.2023 & 6.67 & 3.33 & 0.000 \\
MoT-FS & 0.2687 & 0.5030 & 0.3290 & 0.1593 & 10.00 & 6.67 & 0.000 \\

\bottomrule
\end{tabularx}

\end{document}